\documentclass[reqno]{amsart}

\usepackage[linesnumbered,ruled,norelsize,vlined,algo2e]{algorithm2e}
\SetKwInput{KwInput}{Input}                
\SetKwInput{KwOutput}{Output}

\usepackage{booktabs}
\usepackage{IEEEtrantools}
\usepackage{latexsym}
\usepackage{graphicx}       
\usepackage{mathrsfs}
\usepackage{hyperref}
\usepackage{dsfont}
\usepackage{caption}

\usepackage[noadjust]{cite}
\usepackage{extarrows}

\usepackage{paralist}
\usepackage{capt-of}
\usepackage{xcolor}
\usepackage{diagbox}
\usepackage{cite}
\usepackage{subfigure}
\usepackage{amsmath,amssymb,amsfonts}
\usepackage{graphicx}
\usepackage{textcomp}
\usepackage{wasysym}
\usepackage{caption}

\usepackage{cite}
\usepackage{color}

\usepackage{cite}
\usepackage{subfigure}
\usepackage{amsmath,amssymb,amsfonts}

\usepackage{graphicx}
\usepackage{textcomp}
\usepackage{wasysym}
\usepackage{caption}
\usepackage{apptools}
\usepackage{color}        
\usepackage{microtype}
\usepackage{overpic}

\usepackage[export]{adjustbox}

\usepackage[normalem]{ulem}

\usepackage{eso-pic}
\usepackage{epsfig}
\usepackage{epstopdf}
\usepackage{mathtools}
\usepackage{lineno}
\usepackage{algorithm}
\usepackage[noend]{algpseudocode}
\allowdisplaybreaks

\usepackage{hyperref}
\usepackage{amsthm}
\usepackage{chngcntr}
\usepackage{enumitem}

\usepackage[justification=justified]{caption}
\allowdisplaybreaks

\usepackage{tikz}
\usepackage[many]{tcolorbox}
\usetikzlibrary{calc}

\usetikzlibrary{positioning,shapes}
\usetikzlibrary{arrows}

\tcbuselibrary{skins}
\newtcolorbox{resp}[1][]{%
enhanced jigsaw,%
colback=gray!5!white,%
colframe=gray!80!black,%
size=small,%
boxrule=1pt,%
halign title=flush center,%
coltitle=black,%
breakable,%
drop shadow=black!50!white,%
attach boxed title to top left={xshift=1cm,yshift=-\tcboxedtitleheight/2,yshifttext=-\tcboxedtitleheight/2},%
minipage boxed title=3cm,%
boxed title style={%
	colback=white,%
	size=fbox,%
	boxrule=1pt,%
	boxsep=2pt,%
	underlay={%
		\coordinate (dotA) at ($(interior.west) + (-0.5pt,0)$);
		\coordinate (dotB) at ($(interior.east) + (0.5pt,0)$);
		\begin{scope}[gray!80!black]
			\fill (dotA) circle (2pt);
			\fill (dotB) circle (2pt);
		\end{scope}
	}%
},%
#1%
}

\newcommand{\R}{{\mathbb{R}}}

\newcommand{\N}{{\mathbb{N}}}

\definecolor{myco}{rgb}{0.55, 0.0, 0.63}

\AtAppendix{\counterwithin{theorem}{section}}

\newtheorem{theorem}{Theorem}[section]

\newtheorem{problem}[theorem]{Problem}

\newtheorem{corollary}{Corollary}

\newtheorem{definition}[theorem]{Definition}
\newtheorem{example}{Example}

\newtheorem{remark}[theorem]{Remark}
\newtheorem{assumption}[theorem]{Assumption}
\numberwithin{equation}{section}   

\begin{document}
	
\begin{abstract}
In this paper, we present the synthesis of secure-by-construction controllers that address safety and security properties simultaneously in cyber-physical systems. Our focus is on studying a specific security property called \emph{opacity}, which characterizes the system's ability to maintain plausible deniability of its secret behavior in the presence of an intruder. These controllers are synthesized based on a concept of so-called (augmented) control barrier functions, which we introduce and discuss in detail. We propose conditions that facilitate the construction of the desired (augmented) control barrier functions and their corresponding secure-by-construction controllers. To compute these functions, we propose an iterative scheme that leverages iterative sum-of-square programming techniques. This approach enables efficient computation of these functions, particularly for polynomial systems. Moreover, we demonstrate the flexibility of our approach by incorporating user-defined cost functions into the construction of secure-by-construction controllers. Finally, we validate the effectiveness of our results through two case studies, illustrating the practical applicability and benefits of our proposed approach.

\end{abstract}

\title[Secure-by-Construction Synthesis for Control Systems]{Secure-by-Construction Synthesis for Control Systems}

\author{Bingzhuo Zhong$^{1}$}
\author{Siyuan Liu$^{2}$}
\author{Marco Caccamo$^{3}$}
\author{Majid Zamani$^{1}$}
\address{$^1$Department of Computer Science, University of Colorado Boulder, USA}
\email{\{bingzhuo.zhong,majid.zamani\}@colorado.edu}
\address{$^2$Division of Decision and Control Systems, KTH Royal Institute of Technology, Stockholm, Sweden}
\email{siyliu@kth.se}
\address{$^3$TUM School of Engineering and Design, Technical University of Munich, Germany}
\email{mcaccamo@tum.de}
\maketitle

\section{Introduction}
Over the past few decades, cyber-physical systems (CPS) have emerged as the technological foundation of an increasingly intelligent and interconnected world. These systems play a crucial role in various domains, ranging from transportation systems and smart grids to medical devices. With their complexity, CPS enable innovative functionalities and offer high-performance capabilities. However, the safety-critical nature of CPS raises significant concerns. Any design faults or malfunctions in these systems can have catastrophic consequences, potentially resulting in loss of life. Given the tight interaction between the cyber components and physical entities within CPS, they are also more susceptible to various security threats and attacks. Consequently, there is a growing need to address both safety and security concerns in modern CPS. Ensuring the reliable operation of these systems across different domains has become imperative. Several studies, such as those in \cite{sandberg2015cyberphysical,liu2022secure,dibaji2019systems}, emphasize the importance of tackling these issues and provide valuable insights into secure CPS design and operation. 

Various formal verification and synthesis techniques have been investigated to ensure safety in CPS \cite{knight2002safety,ames2016control,tabuada2009verification,belta2017formal}. Abstraction-based methods have gained significant popularity in the last two decades for safety analysis of CPS \cite{tabuada2009verification,belta2017formal,Zhong2021Safe,Zhong2023Automata,lavaei2022automated}. These methods approximate original systems with continuous state and input sets by their finite abstractions, constructed by discretizing the original sets. Unfortunately, this approach often encounters the ``curse of dimensionality,” resulting in exponential computational complexity growth with system dimension. Alternatively, set-based approaches \cite{Rungger2017Computing,Blanchini2015Set,zhong2021formal,Liu2019Compositional} or (control) barrier functions \cite{prajna2004safety,ames2019control,Nejati2022Data,Choi2021Robust,Jankovic2018Control} can be utilized for verification and controller synthesis without requiring finite abstractions. These approaches ensure overall CPS safety. In summary, CPS design has increasingly prioritized safety. However, security analysis is often postponed to later stages, leading to expensive and time-consuming validation procedures. These challenges motivate researchers \cite{liu2022secure} to develop an integrated approach that addresses safety and security issues \emph{simultaneously} for CPS, leveraging the principles of \emph{correct-by-construction} synthesis \cite{belta2017formal}.

The primary objective of this paper is to propose an approach for synthesizing \emph{secure-by-construction} controllers in safety- and security-critical CPS. To achieve this, we expand the existing correct-by-construction controller synthesis paradigm to encompass security requirements alongside safety considerations, all within a unified framework. To begin exploring the construction of this comprehensive framework, we focus on a specific class of security properties closely tied to CPS information flow, known as ``\emph{opacity}" \cite{lafortune2018history,hadjicostis2020estimation}. Opacity, which is a confidentiality property \cite{dibaji2019systems}, ensures that a system's secret behaviors remain plausibly deniable in the presence of malicious external observers, also referred to as intruders. In this context, a system is deemed opaque if it is impossible for an external intruder to uncover the system's secrets through the information flow. 

The concept of opacity was initially introduced in \cite{mazare2004using} as a means to analyze the performance of cryptographic protocols. Since then, various opacity notions have been proposed to capture different security requirements and information structures in discrete-event systems (DES) (see, for example \cite{lafortune2018history,hadjicostis2020estimation,balun2021comparing}). Among the most widely adopted notions of opacity are language-based opacity \cite{lin2011opacity} and state-based opacity \cite{saboori2007notions}, which includes variants such as initial-state opacity \cite{saboori2013verification}, current-state opacity \cite{saboori2007notions}, infinite-step opacity \cite{saboori2012verification}, and pre-opacity \cite{yang22}. Building upon these notions, numerous works have been developed for the verification or synthesis of controllers with respect to opacity in DES modeled as finite-state automata \cite{saboori2012verification,saboori2013verification,yin2017new} or Petri nets \cite{tong2022verification}.

\textbf{Related work.}
Over the past years, there has been a growing interest in analyzing opacity for \emph{continuous-space} control systems \cite{ramasubramanian2020notions,an2020opacity,yin2020approximate,liu2020verification,liu2021compositional,hou23,liu2021verification,liu2022secure,mizoguchi2021abstraction}. Concretely, \cite{yin2020approximate} extended the concept of opacity in discrete-event systems (DES) to continuous-space cyber-physical systems (CPS) by introducing the notion of approximate opacity for metric systems, which takes into account the imprecision of external observations. Subsequently, various methods were developed to verify whether a given control system is approximately opaque \cite{yin2020approximate,liu2020verification,liu2021compositional,kalat2021modular}. However, only a few recent results \cite{hou2019abstraction,an2020opacity,mizoguchi2021abstraction,Xie2021Secure} address the problem of controller synthesis for opacity properties over CPS. Among these works, \cite{hou2019abstraction} proposed an abstraction-based controller synthesis approach for CPS with finite state sets, leveraging opacity-preserving alternating simulation relations between the original systems and their abstractions. 

The results presented in \cite{Xie2021Secure} addressed the controller synthesis problem for finite Markov decision processes that model stochastic systems with security considerations. The results in \cite{an2020opacity} employed a model-free approximation-based Q-learning method to tackle the opacity enforcement problem in linear discrete-time control systems. It is important to note that all the aforementioned results focus on verification or controller synthesis with respect to opacity, without taking safety considerations into account. More recently, \cite{mizoguchi2021abstraction} proposed a two-stage controller synthesis scheme to enforce safety and approximate opacity in control systems with finite output sets. Specifically, they first synthesized an abstraction-based safety controller without considering opacity properties. In the second stage, control inputs and state transitions that violate the approximate opacity of the system are eliminated. It should be noted that this two-stage approach can lead to an overly conservative controller since opacity is addressed in the later stage, potentially imposing excessive restrictions on the safety controller obtained in the first stage.

\textbf{Our contribution.}
In this paper, we present an \emph{abstraction-free} scheme for constructing secure-by-construction controllers that enforce safety and security properties \emph{simultaneously} in control systems with \emph{continuous} state and input sets. Specifically, we focus on initial-state and infinite-step opacity as the desired security properties, and invariance properties as the safety properties of interest. To synthesize these controllers, we utilize a concept of (augmented) control barrier functions. First, we establish conditions under which (augmented) control barrier functions can be synthesized. We then propose an iterative sum-of-squares (SOS) programming scheme as a systematic approach for computing the desired (augmented) control barrier functions. Additionally, we discuss how user-defined cost functions can be incorporated into the construction of secure-by-construction controllers.

Some of the results presented in this paper have been previously introduced in our preliminary work \cite{Zhong2023Secure}. However, this paper significantly enhances and extends those results in several ways. Firstly, we provide the proofs for all the statements that were omitted in \cite{Zhong2023Secure}. Secondly, we develop secure-by-construction controller synthesis schemes for both initial-state and infinite-step opacity, whereas \cite{Zhong2023Secure} only considered initial-state opacity. Finally, we propose a systematic approach for computing the (augmented) control barrier functions through solving an iterative SOS programming problem, which was not presented in \cite{Zhong2023Secure}.

\section{Problem Formulation} \label{Sec:II} 
\subsection{Notations}
In this paper, we denote by $\R$ and $\N$ the set of real numbers and non-negative integers,  respectively. 
These symbols are annotated with subscripts to restrict them in a usual way, e.g., $\R_{>0}$ denotes the set of positive real numbers. 
For $a,b\in\mathbb{R}$ (resp. $a,b\in\mathbb{N}$) with $a\leq b$, the closed, open and half-open intervals in $\mathbb{R}$ (resp. $\mathbb{N}$) are denoted by $[a,b]$, $(a,b)$, $[a,b)$, and $(a,b]$, respectively. 
Given $N \in \mathbb N_{\ge 1}$ vectors $x_i \!\in\! \mathbb R^{n_i}$, with $i\!\in\! [1;N]$, $n_i\!\in\! \mathbb N_{\ge 1}$, and $n \!= \!\sum_i n_i$, we denote the concatenated vector in $\mathbb R^{n}$ by $x \!=\! [x_1;\!\ldots\!;x_N]$ and the Euclidean norm of $x$ by $\Vert x\Vert$.
Given a set $X$, we denote by $2^X$ the powerset of $X$.
Given a set $Y\subseteq \mathbb{R}^{2n}$, we denote by $\overline{\textbf{Proj}}(Y)$ and $\underline{\textbf{Proj}}(Y)$ the projection of the set $Y$ on to the first and the last $n$ coordinates, respectively, i.e.,
$\overline{\textbf{Proj}}(Y):=\{y\in \mathbb{R}^n | \exists \hat{y}\in\mathbb{R}^n, \text{s.t. } [y;\hat{y}]\in Y\}$, and
$\underline{\textbf{Proj}}(Y): = \{\hat{y}\in \mathbb{R}^n | \exists y\in\mathbb{R}^n, \text{s.t. } [y;\hat{y}]\in Y\}$.
Given a matrix $A$, we denote by 
$A^\top$, 
$\text{trace}(A)$, 
and $\{A\}_{i,j}$, 
the transpose, 
the trace,
and the entry in the $i$-th row and $j$-th column of $A$, respectively.
Given $a_1,\ldots,a_n\in \mathbb{R}$, we denote by $\text{Diag}(a_1,\ldots,a_n)$ a diagonal matrix with $a_1,\ldots,a_n$ on its diagonal.
Given sets $X$ and $Y$, the complement of $X$ w.r.t. $Y$ is defined as $Y \backslash X \!=\! \{x \!\in\! Y \!\mid\! x \!\notin\! X\}.$
Given functions $f\!:\! X \!\rightarrow \!Y$ and $g\!:\! A \!\rightarrow\! B$, we define $f \!\times\! g \!:\! X \!\times\! A \!\rightarrow\! Y \!\times\! B$. 

\subsection{Preliminaries}
First, we recall some definitions from~\cite[Section 3]{Chesi2010LMI}, which are required throughout this paper.
\begin{definition}\label{def:monomials}
	(\emph{Monomial} and \emph{matrix monomial})
	Consider $x:=[x_1;\ldots;x_n]\in\mathbb{R}^n$.
	A \emph{monomial} $m:\mathbb{R}^n\rightarrow \mathbb{R}$ in $x$ is a function defined as $m(x):=x_1^{\alpha_1}x_2^{\alpha_2}\cdots x_n^{\alpha_n}$, with $\alpha_1,\ldots,\alpha_n\in \mathbb{N}$, and we denote by $\mathcal{M}(x)$ the sets of all monomials over $x\in\mathbb{R}^n$.
	The \emph{degree} of the monomial $m(x)$ is defined as deg$(m(x)):=\sum_{i=1}^{n}\alpha_i$. 
	Similarly, a function $M:\mathbb{R}^n\rightarrow \mathbb{R}^{r_1\times r_2}$ is a \emph{matrix monomial} if $\{M(x)\}_{i,j}\in\mathcal{M}(x)$, $\forall i\in[1,r_1], \forall j\in[1,r_2]$.
	We denote by $\mathcal{M}^{\mathsf{m}}(x)$ the set of all matrix monomials over $x\in\mathbb{R}^n$.
	Furthermore, the degree of a matrix monomial $M(x)$ is defined as deg$(M(x)):=\max_{i\in[1,r_1],j\in[1,r_2]}\text{deg}(\{M(x)\}_{i,j})$.$\hfill \star$
\end{definition}
\begin{definition}\label{def:polynomials}
	(\emph{Polynomial} and \emph{matrix polynomial})
	A \emph{polynomial} $h:\mathbb{R}^n\rightarrow \mathbb{R}$ of degree $d$ is a sum of a finite number of monomials, as $p(x):=\sum_{i}^{N_{\mathsf{i}}}c_im_i(x)$, with $c_i\in\mathbb{R}$, $m_i(x)\in\mathcal{M}(x)$, and $d:=\max_{c_i\neq 0} \text{deg}(m_i(x))$.
	We denote by $\mathcal{P}(x)$ the set of polynomials over $x\in\mathbb{R}^n$.
	Moreover, a function $\mathbf{P}:\mathbb{R}^n\rightarrow \mathbb{R}^{r_1\times r_2}$ is a \emph{matrix polynomial} if $\{\mathbf{P}(x)\}_{i,j}\in\mathcal{P}(x)$, $\forall i\in[1,r_1], \forall j\in[1,r_2]$.
	We denote by $\mathcal{P}^{\mathsf{m}}(x)$ the set of matrix polynomials over $x\in\mathbb{R}^n$.
	Accordingly, the degree of the matrix polynomial $\mathbf{P}(x)$ is defined as deg$(\mathbf{P}(x)):=\max_{i\in[1,r_1],j\in[1,r_2]}\text{deg}(\{\mathbf{P}(x)\}_{i,j})$.$\hfill \star$
\end{definition}
\begin{definition}
	(\emph{SOS polynomial} and \emph{SOS matrix polynomial})
	A polynomial $p(x)\in \mathcal{P}(x)$ is a \emph{sum-of-square (SOS) polynomial} if there exists $p_1(x),\ldots,p_\mathsf{i}(x)\in\mathcal{P}(x)$ such that $p(x)=\sum_{i=1}^{\mathsf{i}}p_i^2(x)$. 
	Similarly, $\mathcal{P}_S(x)$ denotes the set of all SOS polynomials over $x\in\mathbb{R}^n$.
	Moreover, a matrix polynomial $\mathbf{P}(x)\in \mathcal{P}^{\mathsf{m}}(x)$ is an \emph{SOS matrix polynomial} if there exists $\mathbf{P}_1(x),\ldots,\mathbf{P}_{\mathsf{j}}(x)\in  \mathcal{P}^{\mathsf{m}}(x)$ such that $\mathbf{P}(x)=\sum_{i=1}^{\mathsf{j}}\mathbf{P}_i^{\top}(x)\mathbf{P}_i(x)$.
	The set of SOS matrix polynomials over $x\in\mathbb{R}^n$ is denoted by $\mathcal{P}_S^{\mathsf{m}}(x)$.$\hfill \star$
\end{definition}

In this paper, we focus on discrete-time control systems, as defined below.
\begin{definition}\label{def:sys1}
	A \emph{discrete-time control system} (dt-CS) $\Sigma$ is a tuple $\Sigma:=(X, X_0, U,f, Y,h)$, in which $X\subseteq \mathbb{R}^n$, $X_0 \subseteq X \subseteq \mathbb{R}^n$, $U \subseteq \mathbb{R}^m$, and $Y\subseteq \mathbb{R}^q$ denote the state set, initial state set, input set, and output set, respectively.  
	The function $f:  X\times  U \rightarrow  X$ is the state transition function, and $h: X \rightarrow  Y$ is the output function.$\hfill \star$
\end{definition}
Alternatively, a dt-CS $\Sigma$ can be described by
\begin{align}\label{eq:2}
	\Sigma:\left\{
	\begin{array}{rl}
		x(k+1)=& f(x(k),\nu(k)),\\
		y(k)=&h(x(k)), \quad \quad \quad k\in\mathbb{N},
	\end{array}
	\right.
\end{align}
in which $x(k)\in X$, $\nu(k)\in U$, and $y(k)\in Y$.
We denote by $\nu: = (\nu(0),\ldots,\nu(k),\ldots)$ an input run of $\Sigma$, and by 
$\mathbf{x}_{x_0,\nu}:= (x(0),\ldots,x(k),\ldots)$ a state run of $\Sigma$ starting from initial state $x_0$ under input run $\nu$, i.e., $x(0)=x_0$, $x(k+1)=f(x(k),\nu(k))$, $\forall k\in \mathbb{N}$.
Additionally, given a controller $C:X \rightarrow 2^U$ for the system $\Sigma$, $\nu$ is called \emph{an input run generated by $C$} if $\nu(k)\in C(x(k))$, $\forall k\in \mathbb{N}$. We further denote by $\Sigma_C := \Sigma \times C$  the closed-loop system under the feedback controller $C$. 

In this paper, we focus on designing a secure-by-construction controller for dt-CSs while simultaneously considering security and safety properties in the controller design procedure. 
Here, the safety properties of interest are formally defined below.
	\begin{definition}
		Consider a dt-CS $\Sigma$ as in Definition~\ref{def:sys1}, an \emph{unsafe set} $X_d \subseteq X$, and a controller $C$.
		The closed-loop system $\Sigma_C$ is safe if
		\begin{align}
			\forall x_0\in X_0, \text{ one has } \mathbf{x}_{x_0,\nu}(k)\notin X_d, \forall k \in \mathbb{N},\label{safe_req}
		\end{align}
		where $\nu(k)\in C(x(k))$.$\hfill \star$
\end{definition}
In other words, the desired safety property requires that any state run of the system must not enter the unsafe set.
Meanwhile, the security properties are expressed as an information-flow security property called \emph{opacity}. 
In this context, we assume the existence of an outside observer (a.k.a. \emph{intruder}) that knows the system model.
Without actively affecting the behavior of the system, the intruder aims to infer certain secret information about the system by observing the output sequences remotely.
In this paper, we focus on two important state-based notions of opacity called \emph{approximate initial-state opacity} and \emph{approximate infinite-step opacity}~\cite{yin2020approximate}, which can be used to model security requirements in a variety of applications, including secure cryptographic protocols and tracking problems in sensor networks \cite{saboori2013verification}. 
Here, we define a set $X_s :=\{X_s^{\text{init}},X_s^{\text{inf}}\}$, in which $ X_s^{\text{init}},X_s^{\text{inf}}\subseteq X$ denote the secret state sets related to approximate initial-state and infinite-step opacity, respectively.
In the rest of this paper, we incorporate the unsafe and secret state sets $X_d$ and $X_s$ in the system definition and use $\Sigma=(X, X_0, X_s, X_d,U,f, Y,h)$ to denote a dt-CS under safety and security requirements. 
The formal definitions of approximate initial-state and infinite-step opacity are then recalled from~\cite{yin2020approximate} as follows.
\begin{definition}\label{def:opa}
	Consider a dt-CS $\Sigma\!=\!(X,X_0,X_s, X_d,U,f, Y,h)$, a controller $C$, and a constant $\delta \in \mathbb{R}_{\geq 0}$. 
   The closed-loop system $\Sigma_C$ is said to be
	\begin{itemize}
		\item \emph{$\delta$-approximate initial-state opaque} if for any $x_0 \!\in\! X_0 \cap X^{\text{init}}_s$ and any finite state run $\mathbf{x}_{x_0,\nu}\!=\!(x_0,\dots, x_n)$ generated by $\Sigma_C$, there exists a finite state run $\mathbf{x}_{\hat x_0,\hat \nu}\!=\!(\hat x_0,\dots, \hat x_n)$ generated by $\Sigma$, with $\hat x_0 \!\in\! X_0 \!\setminus\! X^{\text{init}}_s$, s.t.
		\begin{align}
			\Vert h(x_i)-h(\hat x_i)\Vert \leq \delta, \forall i\in[0,n].\label{opa_ini_requirement}
		\end{align}
		\item \emph{$\delta$-approximate infinite-step opaque} if for any $x_0 \!\in\! X_0$, any finite state run $\mathbf{x}_{x_0,\nu}\!=\!(x_0,\dots, x_n)$ generated by $\Sigma_C$, and any $k \in [0,n]$ such that $x_k \in X^{\text{inf}}_s$, there exists a finite state run $\mathbf{x}_{\hat x_0,\hat \nu}\!=\!(\hat x_0,\dots, \hat x_n)$ generated by $\Sigma$, s.t. 
		\begin{align}
			\Vert h(x_i)-h(\hat x_i)\Vert \leq \delta, \forall i\in[0,n],\label{opa_inf_requirement}
		\end{align}
		where $\hat x_0 \!\in\! X_0$ and $\hat x_k \!\in\! X \!\setminus\! X^{\text{inf}}_s$.$\hfill \star$
	\end{itemize}
\end{definition} 
Intuitively, $\delta$-approximate initial-state opacity requires that the intruder is never certain whether the system was initiated from a secret state;
$\delta$-approximate infinite-step opacity requires that the intruder is never certain that whether the system is/was at a secret state for any time instant $k \in \mathbb{N}$.
Here, constant $\delta$ captures the imprecision of the intruder's observation. 
It is also worth noting that the secret state run $\mathbf{x}_{x_0,\nu}$ in Definition~\ref{def:opa} is generated by the closed-loop system $\Sigma_C$, while the non-secrete state run $\mathbf{x}_{\hat x_0,\hat \nu}$ could be generated by the open-loop system $\Sigma$.
This distinction arises because the intruder lacks knowledge of the controller but possesses understanding of the system model.
Additionally, to enforce $\delta$-approximate initial-state (or infinite-step) opacity over $\Sigma$, the secret of the system should at least not be revealed initially; otherwise, both notions of $\delta$-approximate opacity are trivially violated.
Hence, we assume, without loss of generality:
\begin{align} \label{initassum}
	\forall x_0 \!\in\! X_0 \cap X'_s, \{x \!\in\! X | \Vert h(x) - h(x_0)\Vert \!\leq\! \delta \} \nsubseteq X'_s.
\end{align}
with $X'_s\in \{X_s^{\text{init}},X_s^{\text{inf}}\}$.

\vspace{-0.3cm}
\subsection{Main Problem}
Based on all notations and preliminaries above, we are ready to formulate the main problem to be tackled in this paper.
\begin{resp}
	\begin{problem}\label{prob}
		Consider a dt-CS $\Sigma=(X,X_0,X_s,X_d,U, f,Y,h)$ and a constant $\delta \in \mathbb{R}_{\geq 0}$.
		Synthesize a secure-by-construction controller $C:X \rightarrow 2^U$ (if existing) such that the closed-loop system $\Sigma_C$ satisfies:
		\begin{enumerate}
			\item (Safety) $\Sigma_C$ is safe, i.e., ~\eqref{safe_req} holds.
			\item (Security) $\Sigma_C$ is $\delta$-approximate initial-state (resp. infinite-step) opaque as in Definition~\ref{def:opa}.
		\end{enumerate}
	\end{problem}
\end{resp}
Additionally, we deploy the following running example throughout this paper to better illustrate the theoretical results.
\begin{example}
	(Running example) 
	We consider a car moving on a road, as depicted in Figure~\ref{fig1}.
	A malicious intruder, with observation precision $\delta=0.94$, remotely tracks the car's position.
	Here, the car's initial location, denoted as $X_0$, holds confidential information as it performs a secret task (e.g., transferring money from a bank to an ATM).
	Furthermore, the region $X^{\text{inf}}_s$ represents a special-purpose lane on the road.
	If the intruder confirms the car's entry into this lane, they deduce the car's involvement in a confidential task.
	Our objective is to construct a controller that ensures safety (i.e., keeping the car on the road) while avoiding disclosure of secret information to the intruder (i.e., whether the car is executing a confidential task or not). 
	\begin{figure}[t!]
		\vspace{0.15cm}
		\centerline{			\includegraphics[width=0.5\textwidth]{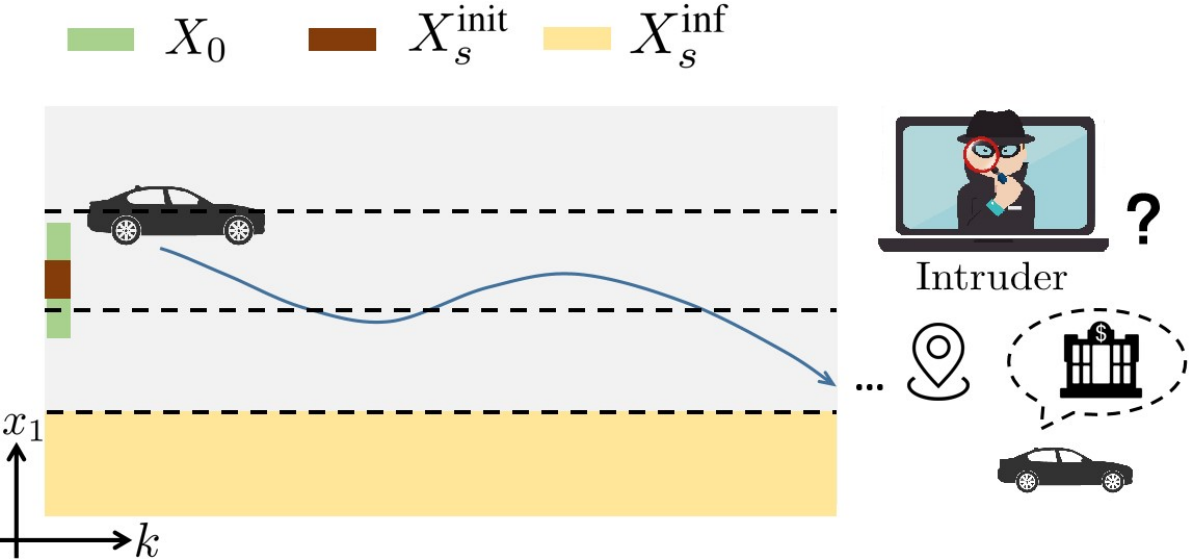}		}
		\caption{A malicious intruder observes the car's position to identify whether or not the car is executing a confidential task.}	\label{fig1}\vspace{-0.5cm}
	\end{figure}
	
	In the running example, we consider a car modeled by
	\begin{align*} 
		\begin{bmatrix}
			x_1(k+1)\\x_2(k+1)
		\end{bmatrix}	
		&= \begin{bmatrix}
			1& \Delta \tau\\
			0& 1
		\end{bmatrix}\begin{bmatrix}
			x_1(k)\\x_2(k)
		\end{bmatrix}+\begin{bmatrix}
			\Delta \tau^2/2\\ \Delta \tau
		\end{bmatrix}\nu(k),\nonumber\\   
		y(k) &= x_1(k),
	\end{align*} 
	where $x_1$ and $x_2$ are the absolute vertical position and velocity of the car in the road frame, respectively, which are perpendicular to the road; 
	$u\in[-4,4] \mathrm{m/s^2}$ is the vertical acceleration of the car as the control input;
	$\Delta \tau=0.1s$ is the sampling time;
	and $y$ is the output of the system which is observed by a malicious intruder.
	Here, we consider the
	state set $X := [-10, 10] \times [-6,6]$,
	initial set $X_0 := [2.7, 3.3] \times \{0\}$,
	secret sets $X^{\text{init}}_s:= [2.8, 3.2] \times \{0\}$ and $X^{\text{inf}}_s:= [5.57, 7.5] \times [-5, 5]$,
	and unsafe set $X_d:=\{[-10,-6.5)\cup(7.5,10]\}\times \{[-6,-5)\cup(5,6]\}$.
	In straightforward terms, it is necessary for the car to remain within the range of $[-6.5,7.5]$ and not exceed an absolute vertical velocity of $5$ m/s to ensure safety.
	Additionally, the desired security requirement can be described using the concepts of $\delta$-approximate initial-state and infinite-step opacity, as defined in Definition~\ref{def:opa}, where $\delta$ is set to $0.94$. 
	$\hfill\diamond$
\end{example}

\section{Synthesis of Secure-by-Construction Controllers}\label{sec3}
In this section, we discuss how to construct secure-by-construction controllers as introduced in Problem~\ref{prob}. 
Concretely, we first propose in Section~\ref{sec3.1} notions of (augmented) control barrier functions for enforcing both safety and opacity properties.
Leveraging these functions, we then discuss in Section~\ref{implementation} how to design secure-by-construction controllers by solving a quadratic program (QP) considering some user-defined cost functions.

\subsection{Control Barrier Functions for Secure-by-Construction Controller Synthesis}\label{sec3.1}

Consider a dt-CS $\Sigma\!=\!(X,X_0,X_s,X_d,U,f,Y,h)$ as in Definition \ref{def:sys1}. 
To tackle Problem~\ref{prob}, we introduce an augmented system associated with $\Sigma$, which is the product between $\Sigma$ and itself, defined as
\begin{align} 
	\Sigma \times \Sigma = (X & \!\times\! X, X_0  \!\times\! X_0, X_s \!\times X_s,X_d \!\times\! X_d,U \!\times\! U,f \!\times\! f, Y  \!\times\! Y, h \!\times\! h).\label{aug_sys}
\end{align}
Here, we denote by $(x,\hat x) \!\in\! X \!\times\! X$ a state pair of  $\Sigma \!\times \!\Sigma$, and by $(\mathbf{x}_{x_0,\nu},  \mathbf{x}_{\hat x_0,\hat \nu})$ the state trajectory of $\Sigma \times \Sigma$ starting from $(x_0, \hat x_0)$ under input run ($\nu, \hat \nu$).
Moreover, we use $\mathcal{R}\!=\!X \!\times X$ to represent the augmented state set.
Having the augmented system, we show that one can synthesize controllers to enforce both safety and security properties over dt-CS by leveraging a notion of (augmented) control barrier functions. 
To this end, the following definition is required.
\begin{definition}\label{cond}
	Consider a dt-CS $\Sigma\!=\!(X,X_0,X_s,X_d,U,f,Y,h)$ as in Definition \ref{def:sys1}.
	Given some sets $\mathcal{R}_0, \mathcal{R}_d\subset \mathcal{R}$, function $\mathcal{B}:X \!\rightarrow\! \mathbb{R}$ is called a \emph{control barrier function} (CBF), and function $\mathcal{B}_O:X \times X \!\rightarrow\! \mathbb{R}$ is called an  \emph{augmented control barrier function}  (ACBF) with respect to $\mathcal{R}_0$ and $ \mathcal{R}_d$,
	if there exists $C:X\rightarrow 2^{U}$ such that the following conditions hold:
	\begin{itemize}
		\setlength{\itemsep}{0pt}
		\setlength{\parsep}{0pt}
		\setlength{\parskip}{0pt}
		\item (\textbf{Cond.1}) $X_0 \subseteq \mathcal{S}$;
		\item (\textbf{Cond.2}) $X_d \subseteq X\backslash \mathcal{S}$;
		\item (\textbf{Cond.3}) $\mathcal{R}_0 \subseteq \mathcal{S}_O$;
		\item (\textbf{Cond.4}) $\mathcal{R}_d \subseteq \mathcal{R} \backslash \mathcal{S}_O$;
		\item (\textbf{Cond.5}) $\forall x \in \mathcal{S}$, $\forall u\in C(x)$, one has $f(x,u)\in\mathcal{S}$;
		\item (\textbf{Cond.6}) $\forall (x,\hat{x}) \in \mathcal{S}_O$, $\forall u\in C(x)$, $\exists \hat{u}\in U$, such that one has $\big(f(x,u),f(\hat{x},\hat{u})\big)\in\mathcal{S}_O$;
	\end{itemize}
	where sets $\mathcal{S}$ and $\mathcal{S}_O$ are defined as
	\begin{align}
		\mathcal{S}&:= \{x\in \mathbb{R}^n | \mathcal{B}(x)\leq 0 \};\label{eq:S}\\
		\mathcal{S}_O&:= \{(x,\hat{x})\in \mathbb{R}^n \times \mathbb{R}^n | \mathcal{B}_O(x,\hat{x})\leq 0 \}.\label{eq:So}
	\end{align}
\end{definition}
Intuitively, (\textbf{Cond.1}), (\textbf{Cond.2}), and (\textbf{Cond.5}) require that $\mathcal{S}$ is a controlled invariant set of $\Sigma$ with respect to the controller $C$, which contains $X_0$ and does not intersect with $X_d$.
Similarly, (\textbf{Cond.3}), (\textbf{Cond.4}), and (\textbf{Cond.6}) indicate the existence of a controller $\hat{C}:X\times U \rightarrow 2^{U}$ such that the set $\mathcal{S}_O$ is a controlled invariant set of $\Sigma \times \Sigma$ with respect to a \emph{joint controller $[C;\hat{C}]$}.
This set should encompass $\mathcal{R}_0$ while ensuring no overlap with $\mathcal{R}_d$. 
From now on, sets $\mathcal{S}$ and $\mathcal{S}_O$ are called \emph{CBF-based invariant set} (\emph{CBF-I set}) and \emph{ACBF-based invariant set} (\emph{ACBF-I set}) associated with sets $\mathcal{R}_0$ and $ \mathcal{R}_d$, respectively.
Note that the concrete forms of sets $\mathcal{R}_0$ and $ \mathcal{R}_d$ depend on the opacity properties of interest (cf. ~\eqref{set:isecure_inf},~\eqref{foraexi2},~\eqref{set:iunsafe}, and~\eqref{foraexi1})
More specifically, to synthesize secure-by-construction controllers enforcing both \emph{safety} and \emph{approximate infinite-step opacity}, we define sets: 
\begin{align}
	&\mathcal{R}^{\text{inf}}_{0} \!:=\! \{(x,\hat x) \!\in\! (X_0 \!\setminus\! X^{\text{inf}}_s) \!\times\! X_0 |  ||h(x) \!-\! h(\hat x)|| \!\leq\! \delta\!-\!\epsilon \}\cup   \{(x,\hat x) \!\in\! (X_0 \!\cap\! X^{\text{inf}}_s)\! \times\!\! X_0\!\!\setminus\! X^{\text{inf}}_s |  ||h(x) \!-\! h(\hat x)|| \!\leq\! \delta\!-\!\epsilon \},\!\! \label{set:init_inf}\\ 
		&\!\! \mathcal{R}^{\text{inf}}_{d}  \!:=\! \{(x,\hat x)\!\in \!X^{\text{inf}}_s \!\times\! (X \!\setminus\! X^{\text{inf}}_s) | ||h(x) \!\!-\!\! h(\hat x)|| \!\geq\!\delta \} \cup X^{\text{inf}}_s \times X^{\text{inf}}_s \!\cup \{(x,\hat x) \!\in\! (X \!\setminus\! X^{\text{inf}}_s) \!\times\!X ||h(x) \!\!-\!\! h(\hat x)|| \!\geq\! \delta \}	  \label{set:isecure_inf},
\end{align}
where $\delta \in \mathbb{R}_{\geq 0}$ captures the imprecision of the intruder's observation as in Definition~\ref{def:opa} and $\epsilon\in[0,\delta]$ is an arbitrary real number.
Then, one can deploy the next result to synthesize secure-by-construction controllers enforcing safety and approximate infinite-step opacity.
\begin{resp}
	\begin{theorem} \label{BC_1}
		Consider a dt-CS $\Sigma$ as in Definition \ref{def:sys1}. 
		Suppose that there exist functions $\mathcal{B}:X \!\rightarrow\! \mathbb{R}$, $\mathcal{B}_O:X \times X \!\rightarrow\! \mathbb{R}$, and $C:X\rightarrow 2^{U}$ such that conditions (\textbf{Cond.1}) - (\textbf{Cond.6}) in Definition~\ref{cond} hold, with sets $\mathcal{R}_{d}= \mathcal{R}^{\text{inf}}_{d}$, and $\mathcal{R}_0\neq \emptyset$ such that
		\begin{align}
			\overline{\textbf{Proj}}(\mathcal{R}^{\text{inf}}_0) \subseteq \overline{\textbf{Proj}}(\mathcal{R}_0) , \underline{\textbf{Proj}}(\mathcal{R}_0) \subseteq \underline{\textbf{Proj}}(\mathcal{R}^{\text{inf}}_0), \label{foraexi2}
		\end{align}
		hold, where $\mathcal{R}^{\text{inf}}_{0}$ and $\mathcal{R}^{\text{inf}}_{d}$ are defined as in \eqref{set:init_inf} and \eqref{set:isecure_inf}, respectively.
		Then, $C(x)$ is a secure-by-construction controller that enforces \emph{safety} and approximate \emph{infinite-step} opacity of $\Sigma$ simultaneously.
	\end{theorem}
\end{resp}
The proof of Theorem~\ref{BC_1} is given in the Appendix~\ref{proof}.
In some cases, it may not be easy to find $\mathcal{R}_0$ satisfying~\eqref{foraexi2} even if such $\mathcal{R}_0$ exists.
To solve this issue, we propose a corollary which can be used to construct controllers enforcing safety and approximate infinite-step opacity without requiring the concrete form of the set $\mathcal{R}_0$ satisfying~\eqref{foraexi2}.
\begin{resp}
	\begin{corollary}\label{inf_nocheck}
		Consider a dt-CS $\Sigma$ as in Definition \ref{def:sys1}. Suppose one can find functions $\mathcal{B}:X \!\rightarrow\! \mathbb{R}$, $\mathcal{B}_O:X \times X \!\rightarrow\! \mathbb{R}$, and $C:X\rightarrow 2^{U}$, such that (\textbf{Cond.1})-(\textbf{Cond.2}) and (\textbf{Cond.4})-(\textbf{Cond.6}) in Definition~\ref{cond} hold, with set $\mathcal{R}_d = \mathcal{R}^{\text{inf}}_d$.
		If $\forall x\in \overline{\textbf{Proj}}(\mathcal{R}^{\text{inf}}_0)$, $\exists \hat{x}\in \underline{\textbf{Proj}}(\mathcal{R}^{\text{inf}}_0)$, such that $(x,\hat{x})\in \mathcal{S}_O$, with $\mathcal{R}^{\text{inf}}_0$ as in~\eqref{set:init_inf}, or, equivalently,
		\begin{align}
			\max_{x\in \overline{\textbf{Proj}}(\mathcal{R}^{\text{inf}}_0)}\ \min_{ \hat{x}\in\underline{\textbf{Proj}}(\mathcal{R}^{\text{inf}}_0)} \mathcal{B}_O(x,\hat{x})\leq 0,\label{maxmin2}
		\end{align}
		then controller $C$ enforces safety and approximate infinite-step opacity as in Problem~\ref{prob}.
	\end{corollary}
\end{resp}
The proof of Corollary~\ref{inf_nocheck} is provided in the Appendix~\ref{proof}. 
Note that one may deploy existing results, e.g.~\cite{Lasserre2011Min}, to tackle the \texttt{max-min} problem in~\eqref{maxmin2}. 

Next, we proceed with discussing the design of secure-by-construction controllers enforcing \emph{safety} and \emph{approximate initial-state opacity}.
Given $\delta \in \mathbb{R}_{\geq 0}$, we define sets
\begin{align} 
	\mathcal{R}^{\text{init}}_0:=&\{(x,\hat x) \in (X_0 \cap X^{\text{init}}_s)   \times  X_0 \setminus X^{\text{init}}_s |  \Vert h(x)-h(\hat x)\Vert \leq \delta-\epsilon\},\label{set:init}\\ 
	\mathcal{R}^{\text{init}}_d:=&\{(x,\hat x)\in \mathcal{R} | \!\!\!\! \quad \Vert h(x)-h(\hat x)\Vert \geq \delta \},\label{set:iunsafe}
\end{align}
where $\epsilon\in[0,\delta]$ is any arbitrary real number.
Based on these sets, we propose Theorem~\ref{BC} for synthesizing secure-by-construction controllers.
\begin{resp}
	\begin{theorem} \label{BC}
		Consider a dt-CS $\Sigma$ as in Definition \ref{def:sys1}. 
		Suppose that there exist functions $\mathcal{B}:X \!\rightarrow\! \mathbb{R}$, $\mathcal{B}_O:X \times X \!\rightarrow\! \mathbb{R}$, and $C:X\rightarrow 2^{U}$, such that conditions (\textbf{Cond.1}) - (\textbf{Cond.6}) in Definition~\ref{cond} hold, with sets $\mathcal{R}_d = \mathcal{R}^{\text{init}}_d$, and $\mathcal{R}_0\neq \emptyset$ satisfying
		\begin{align}
			\overline{\textbf{Proj}}(\mathcal{R}^{\text{init}}_0) \subseteq \overline{\textbf{Proj}}(\mathcal{R}_0) , \underline{\textbf{Proj}}(\mathcal{R}_0) \subseteq \underline{\textbf{Proj}}(\mathcal{R}^{\text{init}}_0),\label{foraexi1}
		\end{align}
		in which sets $\mathcal{R}^{\text{init}}_0$ and $\mathcal{R}^{\text{init}}_d$ are defined in \eqref{set:init} and \eqref{set:iunsafe}, respectively.
		Then, $C(x)$ is a secure-by-construction controller 
		that enforces \emph{safety} and approximate \emph{initial-state} opacity of $\Sigma$ simultaneously.
	\end{theorem}
\end{resp}
The proof of Theorem~\ref{BC} is provided in the Appendix~\ref{proof}.
Similar to Corollary~\ref{inf_nocheck}, one can deploy the next corollary to build controllers enforcing safety and initial-state opacity without explicitly having the concrete form of $\mathcal{R}_0$ satisfying~\eqref{foraexi1}.
\begin{resp}
	\begin{corollary}\label{init_nocheck}
		Consider a dt-CS $\Sigma$ as in Definition \ref{def:sys1}. 
		Suppose (\textbf{Cond.1})-(\textbf{Cond.2}) and (\textbf{Cond.4})-(\textbf{Cond.6}) in Definition~\ref{cond} hold for some functions 
		$\mathcal{B}:X \!\rightarrow\! \mathbb{R}$, $\mathcal{B}_O:X \times X \!\rightarrow\! \mathbb{R}$, and $C:X\rightarrow 2^{U}$, with set $\mathcal{R}_d = \mathcal{R}^{\text{init}}_d$. 
		If $\forall x\in \overline{\textbf{Proj}}(\mathcal{R}^{\text{init}}_0)$, $\exists \hat{x}\in \underline{\textbf{Proj}}(\mathcal{R}^{\text{init}}_0)$, such that $(x,\hat{x})\in \mathcal{S}_O$, with $\mathcal{R}^{\text{init}}_0$ as in~\eqref{set:init}, or, equivalently,
		\begin{align}
			\max_{x\in \overline{\textbf{Proj}}(\mathcal{R}^{\text{init}}_0)}\ \min_{ \hat{x}\in\underline{\textbf{Proj}}(\mathcal{R}^{\text{init}}_0)} \mathcal{B}_O(x,\hat{x})\leq 0, \label{maxmin1}
		\end{align}
		then controller $C(x)$ enforces safety and approximate initial-state opacity as in Problem~\ref{prob}.
	\end{corollary}
\end{resp}
For simple presentation, we omit the proof for Corollary~\ref{init_nocheck} since it is similar to that of Corollary~\ref{inf_nocheck}.

\addtocounter{example}{-1}
\begin{example}[continued]
	(Running example)
	With $\delta =0.94$, we construct sets $\mathcal{R}^{\text{init}}_0$ in~\eqref{set:init}, $\mathcal{R}^{\text{init}}_d$ in~\eqref{set:iunsafe}, 
	$\mathcal{R}^{\text{inf}}_0$ in~\eqref{set:init_inf}, and $\mathcal{R}^{\text{inf}}_d$ in~\eqref{set:isecure_inf} as in~\eqref{Rset1}-\eqref{Rset4}, respectively, with $\epsilon:=0.01$.
	\begin{figure*}[ht!]
		\rule[0pt]{\textwidth}{0.05em}
		\begin{align}
			&\mathcal{R}^{\text{init}}_0 \!=\! \{[x_1;x_2] \!\in\! [2.8, 3.2] \!\times\! \{0\},[\hat x_1;\hat x_2] \!\in\! [2.7, 2.8)\cup(3.2, 3.3] \!\times\! \{0\} \mid ||x_1 \!- \!\hat x_1|| \!\leq\! \delta - \epsilon\}\label{Rset1}\\
			&\mathcal{R}^{\text{init}}_d \!=\! \{(x, \hat x) \!\in\! X \!\times\! X \mid ||x_1 \!-\! \hat x_1|| \!\geq\! \delta  \}\label{Rset2}\\
			&\mathcal{R}^{\text{inf}}_0 \!=\! \{[x_1;x_2] \!\in\! [2.7, 3.7] \!\times\! \{0\},[\hat x_1;\hat x_2] \!\in\! [2.7, 3.3] \!\times\! \{0\} \mid ||x_1 \!- \!\hat x_1|| \!\leq\! \delta - \epsilon\}\label{Rset3}\\
			&\mathcal{R}^{\text{inf}}_d \!=\! \{(x_1, x_2) \!\in\! [5.57, 7.5] \!\times\! [-6,6], [\hat x_1;\hat x_2] \!\in\!([-10, 5.57)\cup(7.5,10])\!\times\! [-6,6] \mid ||x_1 \!-\! \hat x_1|| \!\geq\! \delta  \} \cup \{(x_1, x_2) \!\in\! \nonumber\\
			& ([-10,5.57)\cup(7.5,10]) \!\times\! [-6,6], [\hat x_1;\hat x_2] \!\in\!([-10,10])\!\times\!  [-6,6] \mid ||x_1 \!-\! \hat x_1|| \!\geq\! \delta  \} \cup 
			\{(x, \hat x) \!\in\! X^{\text{inf}}_s \!\times\! X^{\text{inf}}_s\}\label{Rset4}
		\end{align}
		\rule[0pt]{\textwidth}{0.05em}
	\end{figure*}
	Accordingly, for the desired $\delta$-approximate initial-state opacity, one can select $\mathcal{R}_0$ satisfying~\eqref{foraexi1} as:
	\begin{align}
		&\{[x_1;x_2] \!\in\! [2.8, 3.2] \!\times\! \{0\},[\hat x_1;\hat x_2] \!\in\! [2.75, 2.8)\cup(3.2, 3.25] \!\times\! \{0\} \mid ||x_1 \!- \!\hat x_1|| \!\leq\! \delta - \epsilon\}.\label{R0init}
	\end{align}
	As for the desired $\delta$-approximate infinite-step opacity, one can select $\mathcal{R}_0$ satisfying~\eqref{foraexi2} as:
	\begin{align}
		\{(x,\hat{x})\in X_0 \times X_0 \mid  x = \hat{x}\}. \vspace{0.1cm}
	\end{align}
\end{example}

So far, we have proposed notions of CBF and ACBF for constructing secure-by-construction controllers to enforce both safety and opacity properties. 
	Note that these controllers are in general set-valued maps, i.e., assigning to each $x\in X$ a set of feasible control inputs (see Theorems~\ref{BC_1} and~\ref{BC}).
	To provide a single input at each time instant, instead of randomly selecting a control input within the admissible set, one can introduce a user-defined cost function and compute a single control input at each state by solving a QP.
	In the next section, we explain how to construct secure-by-construction controllers by incorporating user-defined cost functions.

\subsection{Design of Secure-by-Construction Controllers with User-defined Cost Functions}\label{implementation}
Consider a user-defined cost function denoted by $\mathcal{J}: X\times U\rightarrow \mathbb{R}$.
	One can construct secure-by-construction controllers by minimizing such a function using Corollary~\ref{controller}. 
\begin{resp}
	\begin{corollary}\label{controller}
		Consider a dt-CS $\Sigma$, its associated augmented system $\Sigma \times \Sigma$ as in~\eqref{aug_sys}, and a user-defined cost function $\mathcal{J}: X\times U\rightarrow \mathbb{R}$.
		Suppose that one obtains CBF $\mathcal{B}(x)$ and ACBF $\mathcal{B}_O(x,\hat{x})$ by leveraging Theorems~\ref{BC} and~\ref{BC_1}.
		Then, the closed-loop system $\Sigma_C$ is safe and $\delta$-approximate initial-state (resp. infinite-step) opaque as described in Problem~\ref{prob}, with controller $C$ generating control input by solving the following optimization problem at each time step $k\in\mathbb{N}$:
			\begin{align*}
				\textit{OP: }\min_{u\in U} &\ \mathcal{J}(x(k),u) \\
				\mbox{s.t.}\ \ &\mathcal{B}_O(f(x(k),u),f(\hat{x}(k),\hat{u}))\leq 0,  \mathcal{B}(f(x(k),u))\leq 0, \text{ and } \hat{u}\in U, 
			\end{align*}
			with $(x(k), \hat{x}(k))$ being state of the  augmented system $\Sigma \times \Sigma$  at time step $k$.
	\end{corollary}
\end{resp}
\vspace{0.1cm}
The proof of Corollary~\ref{controller} is provided in the Appendix~\ref{proof}.
\begin{figure}
	\begin{center}
		\includegraphics[width=0.35\textwidth]{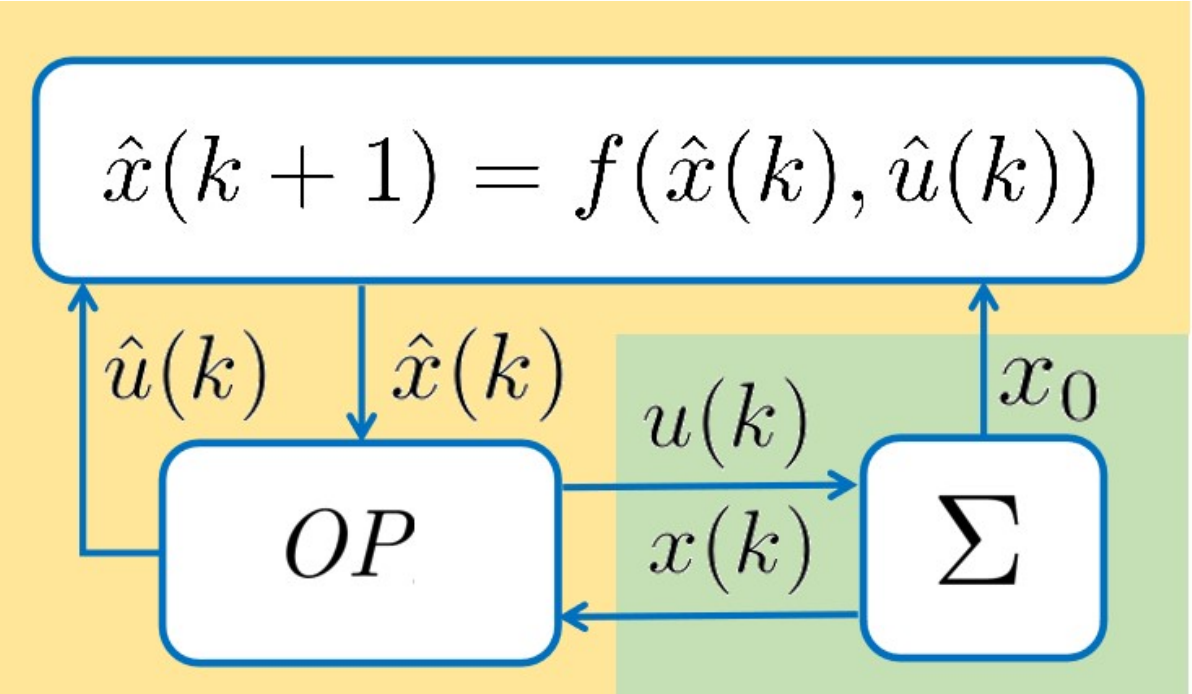}    
		\caption{Secure-by-construction controllers (yellow region) with a user-defined cost function, where $OP$ is an optimization problem appearing in Corollary~\ref{controller}. } \vspace{-0.5cm}
		\label{fig:sbc_controller}
	\end{center}
\end{figure}
Here, the architecture of the controller by leveraging Corollary~\ref{controller} is depicted in Figure~\ref{fig:sbc_controller} and summarized in Algorithm~\ref{alg:implementation}.
	Note that we introduced an internal memory state (denoted by $\hat{x}$ in Figure~\ref{fig:sbc_controller} and Algorithm~\ref{alg:implementation}) for the controller. 
	As a result, one only needs to solve a minimization problem, instead of a \texttt{max-min} problem which is difficult to solve in real-time.

\IncMargin{0.5em}
\begin{algorithm2e}[ht!]
	\DontPrintSemicolon
	\Indm 
	\KwIn{A dt-CS $\Sigma=(X, X_0, X_s, X_d,U,f,Y,h)$ and its associated augmented system $\Sigma \times \Sigma$ as in~\eqref{aug_sys};
		CBF $\mathcal{B}(x)$ and ACBF $\mathcal{B}_O(x,\hat{x})$ as in Theorem \ref{BC}; 
		a user-defined cost function $\mathcal{J}: X\times U\rightarrow \mathbb{R}$;
		and the initial state $x_0$.}
	\KwOut{Control input $u(k)$ at each time step $k\in\mathbb{N}$.}
	\Indp
	$k\gets 0$, $x(0)\gets x_0$.\\
	\While{True}
	{
		\eIf{$k = 0$}{
			Set $\hat{x}(0)\gets\hat{x}_0\!\in\! X_0$, such that $\mathcal{B}_O(x_0,\hat{x}_0)\!\leq\! 0$.
		}
		{
			Update $\hat{x}(k)$ as $\hat{x}(k)\gets f(\hat{x}(k-1),\hat{u}(k-1))$.\\
			Obtain the current state $x(k)$.\\
		}
		Compute $u(k)$ and $\hat{u}(k)$ by solving the optimization problem OP in Corollary~\ref{controller}, with $x:=x(k)$, $\hat{x}:=\hat{x}(k)$.\\
		$k \gets k+1$.\\
	}
	\caption{Running mechanism of secure-by-construction controllers incorporating a user-defined cost function.}
	\label{alg:implementation}
\end{algorithm2e}
\DecMargin{0.5em}

\begin{remark}
	If one only focuses on enforcing approximate initial-state opacity and $x_0\in X_0\backslash X^{\text{init}}_s$, then one can select $\hat{x}_0=x_0$, and set $\hat{\nu}(k)=\nu(k)$ for all $k\in\mathbb{N}$ so that~\eqref{opa_ini_requirement} holds trivially.
	Accordingly, the optimization problem $OP$ in Corollary~\ref{controller} can be reduced to 
	\begin{align*}
		\textit{OP: }\min &\ \mathcal{J}(x,u) \\
		\mbox{s.t.}\ \ & \mathcal{B}(f(x,u))\leq 0, \text{ with } u\in U, x=\mathbf{x}_{x_0,\nu}(k).
	\end{align*}
	Intuitively, if the system does not start from the secrete region $X^{\text{init}}_s$, i.e., $x_0\notin X^{\text{init}}_s$, then any state trajectory $\mathbf{x}_{x_0,\nu}$ always fulfills~\eqref{opa_ini_requirement}.  
	Hence, the conditions for the desired approximate initial-state opacity hold trivially.$\hfill \star$
\end{remark}

Next, we consider again the running example for introducing the desired user-defined cost function for this example.
\addtocounter{example}{-1}
\begin{example}[continued]
	(Running example)
	Here, we consider the following cost function for the running example:
	\begin{align}
		\mathcal{J}(x,u) \!:=\! (x\!-\!x_{\text{set}})^\top \!\!M_1 (x\!-\!x_{\text{set}}) \!+\! 0.1u^2,\label{cost}
	\end{align}
	where $M_1:= \text{Diag}(30, 1)$, and $x_{\text{set}}$ is the desired set point.  $\hfill\diamond$
\end{example}

So far, we have introduced how to construct secure-by-construction controllers (incorporating user-defined cost functions) by leveraging the notions of CBF and ACBF.
In the next section, we focus on the computation of CBF and ACBF over systems with polynomial transition and output functions, and semi-algebraic sets $X_0$, $X_s$, $X_d$, and $X$ (i.e., these sets are described by polynomial equalities and inequalities, cf. Assumption~\ref{assum1}). 
In this case, one can use SOS programming~\cite{JarvisWloszek2005Control} to compute polynomial-type CBF and ACBF leveraging existing semi-definite-programming (SDP) solver (e.g.,~\texttt{Mosek}~\cite{ApS2019MOSEK})).

\section{Iterative Sum-of-square (SOS) Programming for Synthesizing CBF and ACBF}\label{sec:sos_imp}
\subsection{SOS Conditions for Computing CBF and ACBF}\label{sec:sos}
In this section, we focus on computing CBF and ACBF over systems with polynomial transition and output functions, and semi-algebraic sets $X_0$, $X_s$, $X_d$, and $X$.
Here, we formulate these functions and sets in the next assumption.
\begin{assumption}\label{assum1}
	Consider a dt-CS $\Sigma\!=\!(X,X_0,X_s,X_d,U,f,Y,h)$ as in Definition~\ref{def:sys1}.
	We assume:
	\begin{itemize}
		\item $f(x,u)\in \mathcal{P}(x,u)$, $h(x)\in  \mathcal{P}(x)$. \vspace{0.1cm}
		\item The sets $X$, $X_0$, $X_s$, and $X_d$ are defined as 
		\begin{small}
			\begin{align}
				\!\!\!\!\!\!\!\!\!\!\!X \!=\! \bigcup_{\mathsf{x}=1}^{n_\mathsf{x}} X_{\mathsf{x}}, 
				X_0 \!=\! \bigcup_{\mathsf{a}=1}^{n_\mathsf{a}} X_{0,\mathsf{a}}, 
				X_s \!=\! \bigcup_{\mathsf{b}=1}^{n_\mathsf{b}} X_{s,\mathsf{b}}, 
				X_d \!=\! \bigcup_{\mathsf{c}=1}^{n_\mathsf{c}} X_{d,\mathsf{c}},\label{eq:poly_set}
			\end{align}
		\end{small}
		where $n_\mathsf{x},n_\mathsf{a},n_\mathsf{b},n_\mathsf{c}\in\mathbb{N}$ are some known integers, and
		\begin{align*}
			X_{\mathsf{x}}&:=\{x\in\mathbb{R}^n\,\big|\,\mu_{\mathsf{x},k}(x)\!\geq\! 0, k \!\in\! [1,\mathsf{k}(\mathsf{x})], \mathsf{k}(\mathsf{x})\!\in\!\mathbb{N}\},\\
			X_{0,\mathsf{a}}&:=\{x\in\mathbb{R}^n\,\big|\,\alpha_{\mathsf{a},k}(x)\!\geq\! 0, k \!\in\! [1,\mathsf{k}_0(\mathsf{a})], \mathsf{k}_0(\mathsf{a})\!\in\!\mathbb{N}\},\\
			X_{s,\mathsf{b}}&:=\{x\in\mathbb{R}^n\,\big|\,\beta_{\mathsf{b},k}(x)\!\geq\! 0, k \!\in\! [1,\mathsf{k}_s(\mathsf{b})], \mathsf{k}_s(\mathsf{b})\!\in\!\mathbb{N}\},\\
			X_{d,\mathsf{c}}&:=\{x\in\mathbb{R}^n\,\big|\,\gamma_{\mathsf{c},k}(x)\!\geq\! 0, k \!\in\! [1,\mathsf{k}_d(\mathsf{c})], \mathsf{k}_d(\mathsf{c})\!\in\!\mathbb{N}\},
		\end{align*}
		with $\mu_{\mathsf{x},k}(x), \alpha_{\mathsf{a},k}(x)$, $\beta_{\mathsf{b},k}(x), \gamma_{\mathsf{c},k}(x) \in \mathcal{P}(x)$ being some known polynomial functions.
		\item The input set $U$ is defined as
		\begin{equation}
			U := \{u\in\mathbb{R}^m\,\big|\,\rho_{j}(u)\leq 0, j = [1,\mathsf{j}] \}\subset \mathbb{R}^m,\label{eq:U}
		\end{equation}
		with $\rho_{j}(u)\in\mathcal{P}(u)$ being some known polynomial functions.$\hfill \star$
	\end{itemize}
\end{assumption}

Based on  Assumption~\ref{assum1}, both $\mathcal{R}^{\text{init}}_d$ in~\eqref{set:iunsafe} and $\mathcal{R}^{\text{inf}}_{d}$ in~\eqref{set:isecure_inf} can be rewritten, without loss of generality, as
	\begin{equation}
 	\medmuskip=1mu
		\thinmuskip=1mu
		\thickmuskip=1mu
		\!\!\!\!\!\!\!\!\!\mathcal{R}_d \!:=  \bigcup_{\mathsf{e}=1}^{n_\mathsf{e}} \{(x,\hat{x})\!\in\!\mathcal{R} \big|\lambda_{\mathsf{e},r}(x,\hat{x})\!\geq\! 0, r \!\in\! [1,\mathsf{r}(\mathsf{e})],\mathsf{r}(\mathsf{e})\!\in\!\mathbb{N}\},\label{Rud}
	\end{equation}
with $n_\mathsf{e}\in\mathbb{N}$ being a known integer, and $\lambda_{\mathsf{e},r}(x,\hat{x}) \in \mathcal{P}(x,\hat{x})$ being some known polynomial functions.
Therefore, for simple presentation, we simply use the notation $\mathcal{R}_d$ in the following discussion and do not distinguish between $\mathcal{R}^{\text{init}}_d$ and $\mathcal{R}^{\text{inf}}_d$ unless necessary.
Similarly, we focus on those $\mathcal{R}_0$ satisfying~\eqref{foraexi1} or~\eqref{foraexi2} being of the form of
	\begin{equation}
 \medmuskip=1mu
		\thinmuskip=1mu
		\thickmuskip=1mu
		\!\!\!\!\!\!\!\!\!\mathcal{R}_0\!:=\! \bigcup_{\!\mathsf{d}=1}^{n_\mathsf{d}} \{(x,\hat{x})\!\in\!\mathcal{R} \big|\kappa_{\mathsf{d},r}(x,\hat{x})\!\geq\! 0,r \!\in\! [1,\mathsf{r}(\mathsf{d})],\mathsf{r}(\mathsf{d})\!\in\!\mathbb{N}\},\label{R0}
	\end{equation}
with $n_\mathsf{d}\in\mathbb{N}$ being a known integer, and $\kappa_{\mathsf{d},r}(x,\hat{x}) \in \mathcal{P}(x,\hat{x})$ being some known polynomial functions.
Additionally, we focus on CBF $\mathcal{B}$ as in~\eqref{eq:S} and ACBF $\mathcal{B}_O$ as in~\eqref{eq:So} in the form of polynomial functions.
Then, one can find $\mathcal{B}$ and $\mathcal{B}_O$ leveraging the next result. 
\begin{resp}
	\begin{theorem}\label{thm:bilinear}
		Consider a dt-CS $\Sigma$ as in Definition~\ref{def:sys1} such that Assumption~\ref{assum1} holds, and a constant $\delta \in\mathbb{R}_{\geq 0}$ as in Definition~\ref{def:opa}.
		If there exists $\mathcal{B}(x),\, k_a(x)\in\mathcal{P}(x)$, and $\mathcal{B}_O(x,\hat{x}),\, \hat{k}_a(x,\hat{x})\in\mathcal{P}(x,\hat{x})$, $\forall a \in[1,m]$ such that~\eqref{cd4}-\eqref{cdd3} hold, then functions $\mathcal{B}(x), \mathcal{B}_O(x,\hat{x})$ are CBF and ACBF as in~\eqref{eq:S} and~\eqref{eq:So}, respectively.
	\end{theorem}
\end{resp} 
The proof of Theorem~\ref{thm:bilinear} can be found in the Appendix~\ref{proof_imp}.
Intuitively, conditions~\eqref{cd4}-\eqref{cdd2} correspond to (\textbf{Cond.1}) - (\textbf{Cond.6}) in Definition~\ref{cond}, respectively, while~\eqref{cd3} and~\eqref{cdd3} ensure the existence of $u,\hat{u}\in U$ in (\textbf{Cond.5)} and (\textbf{Cond.6}) in Definition~\ref{cond}, respectively, with $u=[k_1(x);\ldots;k_m(x)]$, and $\hat{u}=[\hat{k}_1(x,\hat{x});\ldots;\hat{k}_m(x,\hat{x})]$.
So far, we have proposed SOS conditions under which CBF and ACBF exist for a given dt-CS. 
	Next, we proceed with discussing how to compute CBF and ACBF systematically via an iterative scheme by leveraging these conditions.

\subsection{Iterative Scheme for Computing CBF and ACBF}\label{sec:iterative_approach}
One may notice that constraints~\eqref{cd4}-\eqref{cdd3} are bilinear between functions $\mathcal{B}(x)$, $\mathcal{B}_O(x,\hat{x})$, $k_a(x)$, $\hat{k}_a(x,\hat{x})$, with $a\in[1,m]$, and those unknown (SOS) polynomial multipliers.
Here, we introduce an iterative scheme to compute these functions over dt-CS that are polynomial control-affine systems~\cite{khalil2002nonlinear}, described as
\begin{align}
	x(k+1)\!:=\!  A\mathcal{H}(x(k))x(k)\!+\!B\mathcal{U}(x(k))\nu(k),k\!\in\!\mathbb{N},\label{sys2}
\end{align}
where $A \in \mathbb{R}^{n\times N_x}$ and $B \in \mathbb{R}^{n\times N_u}$ are some known constant matrices, and $\mathcal{U}(x),\mathcal{H}(x)\in \mathcal{M}^{\mathsf{m}}(x)$ are some known matrix monomials with appropriate dimensions.
Concretely, in the proposed iterative scheme, we will first compute an \emph{initial CBF-I set}, denoted by $\mathcal{S}^{\text{init}}$, as well as an \emph{initial ACBF-I set}, denoted by $\mathcal{S}_O^{\text{init}}$, such that (\textbf{Cond.2}), (\textbf{Cond.4}), (\textbf{Cond.5}), and (\textbf{Cond.6}) in Definition~\ref{cond} are satisfied (these conditions correspond to~\eqref{cd1}, and~\eqref{cdd1}-\eqref{cdd3}, respectively).
Then, we will propose an iterative scheme to expand the regions characterized by $\mathcal{S}^{\text{init}}$ and $\mathcal{S}_O^{\text{init}}$.
In each iteration, we will check whether or not~\eqref{cd4} (resp.~\eqref{maxmin1}) and~\eqref{cdd4} (resp.~\eqref{maxmin2}) hold
over the expanded version of CBF-I set (referred to as \emph{expanded CBF-I set} and denoted by $\mathcal{S}^{\text{exp}}$) and of ACBF-I set (referred to as \emph{expanded ACBF-I set} and denoted by $\mathcal{S}_O^{\text{exp}}$).

\subsubsection{Computation of Initial CBF-I and ACBF-I Sets}
To compute the initial CBF-I set $\mathcal{S}^{\text{init}}$ and the initial ACBF-I set $\mathcal{S}_O^{\text{init}}$ over the system as in~\eqref{sys2}, one first selects sets
\begin{align}
	\bar{X} &:= \{x\in X| a_ix\leq 1, i \in [1,\mathsf{i}] \}\subseteq X\backslash X_d\label{X_sub},\\
	\bar{\mathcal{R}} &:= \{(x,\hat{x})\in \mathcal{R}|b_t[x;\hat{x}]\leq 1,t \in [1,\mathsf{t}]\}\subseteq \mathcal{R}\backslash \mathcal{R}_d, \label{R_sub}
\end{align}
with $a_i\in \mathbb{R}^n$, and $b_t\in \mathbb{R}^{2n}$ being some known constant vectors, $X$ and $X_d$ being as in~\eqref{eq:poly_set}, $\mathcal{R}_d$ being as in~\eqref{Rud}, and $\mathcal{R}$ being the state set of $\Sigma\times\Sigma$.
Additionally, we propose some conditions that are required for computing $\mathcal{S}^{\text{init}}$ and $\mathcal{S}_O^{\text{init}}$.
\begin{definition}\label{def:some_cond}
	Consider a polynomial system as in~\eqref{sys2}, and sets $\bar{X}$, $\bar{\mathcal{R}}$ as in~\eqref{X_sub} and~\eqref{R_sub}, respectively. 
	We define the following conditions: 
	\begin{align}
		&\begin{bmatrix} Q& \mathsf{g}(x)^\top \\\mathsf{g}(x) &Q\end{bmatrix}\in \mathcal{P}_S^{\mathsf{m}}(x),\label{init_cond1}\\
		&a_iQa^\top_i\leq 1,\,i\in[1,\mathsf{i}];\label{init_cond1-1}\\
		&\begin{bmatrix} Q_o& \mathsf{g}_o(x,\hat{x})^\top \\\mathsf{g}_o(x,\hat{x}) &Q_o\end{bmatrix}\in \mathcal{P}_S^{\mathsf{m}}(x,\hat{x}),\label{init_cond2}\\
		&b_tQ_ob^\top_t\leq 1,\,t\in[1,\mathsf{t}];\label{init_cond2-1}
	\end{align}
	for some $Q\in \mathbb{R}^{n \times n}$, $Q_o\in \mathbb{R}^{2n \times 2n}$, $\bar{K}(x)\in\mathcal{P}^{\mathsf{m}}(x)$, and $\bar{K}_o(x,\hat{x})\in\mathcal{P}^{\mathsf{m}}(x,\hat{x})$, 
	with $\mathsf{g}(x)\!:=\!A\mathcal{H}(x)Q+B\mathcal{U}(x)\bar{K}(x)$, $\mathsf{g}_o(x,\hat{x})\!:=\!A_o(x,\hat{x})Q_o+B_o(\hat{x})\bar{K}'_o(x,\hat{x})$, where $\mathcal{H}(x)\in\mathcal{P}(x)$ is as in~\eqref{sys2}, 
	\begin{small}
		\begin{align*}
			A_o:=\begin{bmatrix}A\mathcal{H}(x)+B\mathcal{U}(x)K(x)& \mathbf{0}\\\mathbf{0}&A\mathcal{H}(\hat{x})\end{bmatrix},
			B_o:=\begin{bmatrix}\mathbf{0}& \mathbf{0}\\\mathbf{0}&B\mathcal{U}(\hat{x})\end{bmatrix},
		\end{align*}
	\end{small}
	$\!\!\bar{K}'_o(x,\hat{x}):=[\mathbf{0};\bar{K}_o(x,\hat{x})]$, $K(x):=\bar{K}(x)Q^{-1}$, and $\mathbf{0}$ are zero matrices with appropriate dimensions.$\hfill \star$
\end{definition}

Note that one can check conditions~\eqref{init_cond1}-\eqref{init_cond2-1} using semi-definite-programming (SDP) solver (e.g.,~\texttt{Mosek}~\cite{ApS2019MOSEK}).
With Definition~\ref{def:some_cond}, the next result shows how to compute the sets $\mathcal{S}^{\text{init}}$ and $\mathcal{S}_O^{\text{init}}$ over the system as in~\eqref{sys2}. 
\begin{resp}
	\begin{theorem}\label{opt_nonlinear_init}
		Suppose that there exists positive-definite matrices $Q\in \mathbb{R}^{n \times n}$, $Q_o\in \mathbb{R}^{2n \times 2n}$, and matrix polynomials $\bar{K}(x)\in\mathcal{P}^{\mathsf{m}}(x),\bar{K}_o(x,\hat{x})\in\mathcal{P}^{\mathsf{m}}(x,\hat{x})$ such that conditions~\eqref{init_cond1}-\eqref{init_cond2-1} hold.
		Then, there exists $c_1,c_2\in(0,1]$ such that (\textbf{Cond.2}), (\textbf{Cond.4}), (\textbf{Cond.5}), and (\textbf{Cond.6}) in Definition~\ref{cond} hold, with 
		\begin{align}
			\!\!\!\mathcal{S}&:= \{x\in \mathbb{R}^n | x^\top Q^{-1}x-c_1\leq 0\},\label{eq1}\\
			\!\!\!\mathcal{S}_O&\!:=\! \{(x,\hat{x})\!\in\! \mathbb{R}^n \!\times\! \mathbb{R}^n | [x;\hat{x}]^\top\!\! Q_o^{-1}[x;\hat{x}]\!-\!c_2\!\leq\! 0 \}.\label{eq2}
		\end{align}
	\end{theorem}
\end{resp}
The proof of Theorem~\ref{opt_nonlinear_init} can be found in the Appendix~\ref{proof_imp}.
Having Theorem~\ref{opt_nonlinear_init}, one obtains the initial CBF-I set $\mathcal{S}^{\text{init}}$ as in~\eqref{eq1} and initial ACBF-I set $\mathcal{S}_O^{\text{init}}$ as in~\eqref{eq2} by deploying Algorithm~\ref{alg:initial} in Appendix~\ref{ap3}.
As a key insight, in Algorithm~\ref{alg:initial}, step~\ref{step_1} aims at computing $Q$ and $\bar{K}(x)$ in~\eqref{init_cond1}; step~\ref{step_2} aims at computing $Q_o$ and $\bar{K}_o(x,\hat{x})$ in~\eqref{init_cond2}; steps~\ref{step_3} and~\ref{step_4} are for computing $c_1$ and $c_2$ to respect the input constraints as in~\eqref{eq:U}. 
Next, we revisit the running example to show how to compute sets $\mathcal{S}^{\text{init}}$ and $\mathcal{S}_O^{\text{init}}$ leveraging Theorem~\ref{opt_nonlinear_init} and Algorithm~\ref{alg:initial}.
\addtocounter{example}{-1}
\begin{example}[continued]
	(Running example)
	To compute the sets $\mathcal{S}^{\text{init}}$ and $\mathcal{S}_O^{\text{init}}$ for the running example, we select 
	\begin{itemize}
		\item the set $\bar{X}$ as in~\eqref{X_sub}, with $a_1:=[-0.1554\ 0]$, $a_2:=[0.1347;0]^\top$, $a_3:=[0;-0.2020]^\top$, and $a_4:=-a_3$;
		\item the set $\bar{\mathcal{R}}$ as in~\eqref{R_sub}, with $b_1 := [1.086;0;-1.086;0]^\top$, $b_2 := -b_1$, and $b_3 := [0;0;0.1813;0]^\top$;
		\item candidates of $\bar{K}(x)$ and $\bar{K}_o(x,\hat{x})$ as in Theorem~\ref{opt_nonlinear_init}, with deg$(\bar{K}(x))=0$ and deg$(\bar{K}_o(x,\hat{x}))=0$.
	\end{itemize}
	Then, we deploy Theorem~\ref{opt_nonlinear_init} and Algorithm~\ref{alg:initial} to compute $\mathcal{S}^{\text{init}}$ and $\mathcal{S}_O^{\text{init}}$.
	Accordingly, we obtain 
	\begin{align*}
		Q &= \begin{bmatrix} 40.155 &-8.950\\-8.950 &19.845\end{bmatrix}\!,\ \bar{K} = \begin{bmatrix} -0.014 &-0.033\end{bmatrix}\!,\\
		Q_o &= \begin{bmatrix} 35.612 &-13.132&32.277&-16.008\\-13.132 &25.424& -11.908 & 26.485\\32.277 &-11.908 &29.520 &-14.952\\-16.008 &26.485 &-14.952 &29.445 \end{bmatrix}\!,\text{ and }	\bar{K}_o = \begin{bmatrix}48.675&16.387&-54.999&-16.276\end{bmatrix},
	\end{align*}
	with $c_1=1$, and $c_2= 0.57$. $\hfill\diamond$
\end{example}

\subsubsection{Iterative Scheme for Computing the Expanded CBF-I and ACBF-I Sets}\label{sec:iterative}

\IncMargin{0.5em}
\begin{algorithm2e}[ht!]
	\DontPrintSemicolon
	\Indm 
	\KwIn{Sets $\mathcal{S}^{\text{init}}$ and $\mathcal{S}_O^{\text{init}}$, matrices $Q$ and $Q_o$, and matrix polynomials $\bar{K}(x)\in\mathcal{P}^{\mathsf{m}}(x),\bar{K}_o(x,\hat{x})\in\mathcal{P}^{\mathsf{m}}(x,\hat{x})$ obtained by leveraging Theorem~\ref{opt_nonlinear_init} and Algorithm \ref{alg:initial}.
		Degrees of $\mathcal{Z}(x)$ in~\eqref{quad:B}, $\mathcal{Z}_o(x,\hat{x})$ in~\eqref{quad:Bo}, and those (SOS) polynomial multipliers appearing in~\eqref{cd4}-\eqref{cdd3};
		maximal number of iterations $i_{\text{max}}$;
		constant $\lambda\in\mathbb{R}_{\geq 0}$ selected by users. }
	\KwOut{Expanded CBF-I set $\mathcal{S}^{\text{exp}}$ and ACBF-I set $\mathcal{S}^{\text{exp}}_O$.}
	\Indp	
	$i\gets 1$, $[k_1(x);\ldots;k_m(x)]\gets \bar{K}(x)Q^{-1}x$ and $[\hat{k}_1(x,\hat{x}); \ldots;\hat{k}_m(x,\hat{x})]\gets \bar{K}_o(x,\hat{x})Q_o^{-1}[x;\hat{x}]$.\\
	\While{$i\leq i_{\text{max}}$}
	{
		Fix $\mathcal{B}(x)$, $\mathcal{B}_O(x,\hat{x})$, $k_a(x)$, and $\hat{k}_a(x,\hat{x})$, with $a\in[1,m]$, compute the (SOS) polynomial multipliers in~\eqref{cd1}, and~\eqref{cdd1}-\eqref{cdd3}\label{fix1}.\\
		Fix the (SOS) polynomial multipliers in~\eqref{cd1}, and~\eqref{cdd1}-\eqref{cdd3} obtained in step~\ref{fix1}, solve \textit{OP$_1$} to compute new $\mathcal{B}(x)$, $\mathcal{B}_O(x,\hat{x})$, $k_a(x)$, and $\hat{k}_a(x,\hat{x})$, with $a\in[1,m]$:\label{fix2}
		\begin{align}
			\textit{OP$_1$}:\min  &\  \text{trace}(P)+\lambda\text{trace}(P_O) \label{obj_f}\\
			\mbox{s.t.}&\ \text{\eqref{cd1}, and~\eqref{cdd1}-\eqref{cdd3} hold}.\nonumber
		\end{align}\\
		Check whether~\eqref{cd4} (resp.~\eqref{maxmin1}) is feasible.\label{fea1}\\
		Check whether~\eqref{cdd4} (resp.~\eqref{maxmin2}) is feasible.\label{fea2}\\
		\If{steps~\ref{fea1} and~\ref{fea2} are feasible}{
			Stop successfully.
		}
		\eIf {$i > i_{\text{max}}$}{
			Stop inconclusively.
		}
		{
			$i\gets i+1$.
		}
	}
	\caption{Iterative scheme for computing CBF and ACBF.}\label{alg:iterative}
\end{algorithm2e}
\DecMargin{0.5em}

In this subsection, we proceed with discussing how to compute the expanded CBF-I set $\mathcal{S}^{\text{exp}}$ and ABCF-I set $\mathcal{S}_O^{\text{exp}}$ using Theorem~\ref{thm:bilinear} based on the sets $\mathcal{S}^{\text{init}}$ and 
$\mathcal{S}_O^{\text{init}}$ obtained by leveraging Theorem~\ref{opt_nonlinear_init}.
Since we focus on polynomial type CBF and ACBF, without loss of generality~\cite[Section 1.2]{Chesi2011Domain}, one can write $\mathcal{B}$ and $\mathcal{B}_O$ in Definition~\ref{cond} as 
\begin{align}
	\mathcal{B}(x) &:= \mathcal{Z}^\top(x)P\mathcal{Z}(x), \label{quad:B}\\
	\mathcal{B}_O(x,\hat{x})&:= \mathcal{Z}_o^\top(x,\hat{x})P_o\mathcal{Z}_o(x,\hat{x}),\label{quad:Bo}
\end{align}
respectively, in which $\mathcal{Z}(x)\in\mathcal{P}^{\mathsf{m}}(x)$ and $\mathcal{Z}_o(x,\hat{x})\in \mathcal{P}^{\mathsf{m}}(x,\hat{x})$, with $P$ and $P_o$ being real matrices with appropriate dimensions.
Then, one can deploy Algorithm~\ref{alg:iterative} to compute the sets $\mathcal{S}^{\text{exp}}$ and $\mathcal{S}_O^{\text{exp}}$ according to Theorem~\ref{thm:bilinear}. 
If Algorithm~\ref{alg:iterative} does not stop successfully, one may consider selecting larger iteration number $i_{\text{max}}$, or increasing the degrees of $\mathcal{Z}(x)$ in~\eqref{quad:B}, $\mathcal{Z}_o(x,\hat{x})$ in~\eqref{quad:Bo}, and those (SOS) polynomial multipliers appearing in~\eqref{cd4}-\eqref{cdd3}.
\begin{remark}
	By deploying Algorithm~\ref{alg:iterative}, one can enlarge the CBF-I and ACBF-I sets characterized by $\mathcal{S}^{\text{init}}$ and $\mathcal{S}_O^{\text{init}}$, respectively, since
	\begin{enumerate}
		\item by deploying the objective function in~\eqref{obj_f}, the volumes of the sets $\mathcal{S}$ and $\mathcal{S}_O$ tend to increase~\cite[Section 4.4.1]{Chesi2011Domain} as the number of iteration increases.
		\item CBF $\mathcal{B}(x)$ and ACBF $\mathcal{B}_O(x,\hat{x})$ in~\eqref{quad:B} and~\eqref{quad:Bo} can be polynomial functions of order higher than 2, while 
			Theorem~\ref{opt_nonlinear_init} only provides polynomial CBF $\mathcal{B}(x)$ and ACBF $\mathcal{B}_O(x,\hat{x})$ of order 2.$\hfill \star$
	\end{enumerate}
\end{remark}

\section{Case Study}\label{sec4}
To show the effectiveness of our results, we first proceed with discussing the running example with the results in Section~\ref{sec:iterative_approach} and simulate the system using the controller proposed in Section~\ref{implementation}.
Then, we apply our results to a case study on controlling a satellite.

\subsection{Running Example (continued)}\label{cases:car}
Based on the initial CBF-I set $\mathcal{S}^{\text{init}}$ and the initial ACBF-I set $\mathcal{S}_O^{\text{init}}$ obtained by leveraging Theorem~\ref{opt_nonlinear_init}, we deploy the iterative scheme proposed in Section~\ref{sec:iterative} to compute the expanded CBF-I set $\mathcal{S}^{\text{exp}}$ and ACBF-I set $\mathcal{S}_O^{\text{exp}}$.
Here, we consider candidates of $\mathcal{B}(x)$, $\mathcal{B}_O(x,\hat{x})$, $k(x)$, and $\hat{k}(x,\hat{x})$ with deg$(\mathcal{B}(x))=4$, deg$(\mathcal{B}_O(x,\hat{x}))=4$, deg$(k(x))=6$, and deg$(\hat{k}(x,\hat{x}))=4$.
The computation ends in 7 iterations, and the evolution of the sets 
$\mathcal{S}^{\text{exp}}$ and $\mathcal{S}_O^{\text{exp}}$ with respect to the number of iterations are depicted in Figure~\ref{fig:CBF} and Figure~\ref{fig:ACBF} in Appendix~\ref{ap3}, respectively.
\begin{figure}
	\begin{center}
		\includegraphics[width=0.45\textwidth]{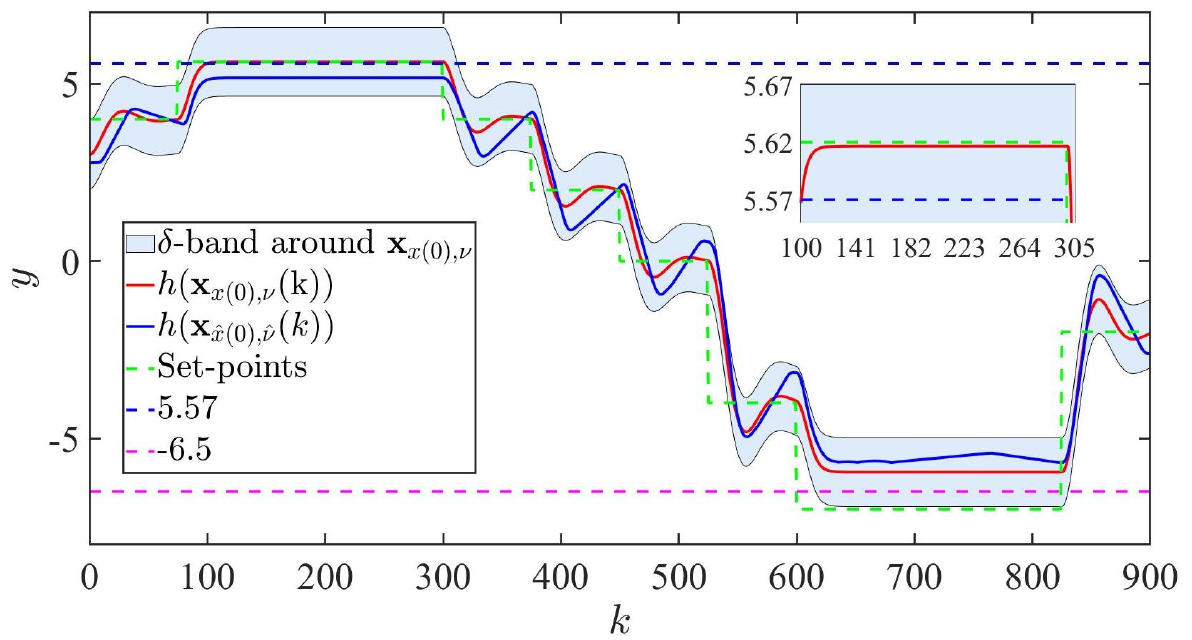}   
		\caption{A state run $\mathbf{x}_{x(0),\nu}$ of the car initiated from a secret location, and its ($\delta$-close output) equivalent  trajectory $\mathbf{x}_{\hat{x}(0),\hat{\nu}}$ started from a non-secret region, i.e., $x(0)\in X_0\cap X^{\text{init}}_s$ and $\hat{x}(0)\in X_0\backslash X^{\text{init}}_s$. The shaded area in light blue is a $\delta$-band around $\mathbf{x}_{x(0),\nu}$. 
			Additionally, $5.57$ and $-6.5$ are the boundaries of the secret state set $X^{\text{inf}}_s$ and the unsafe set $X_d$, respectively.} 
		\label{fig:sim_runningexample}
	\end{center}
\end{figure}
\begin{figure}[ht!]
	\centering
	\subfigure{
		\includegraphics[width=0.5\textwidth]{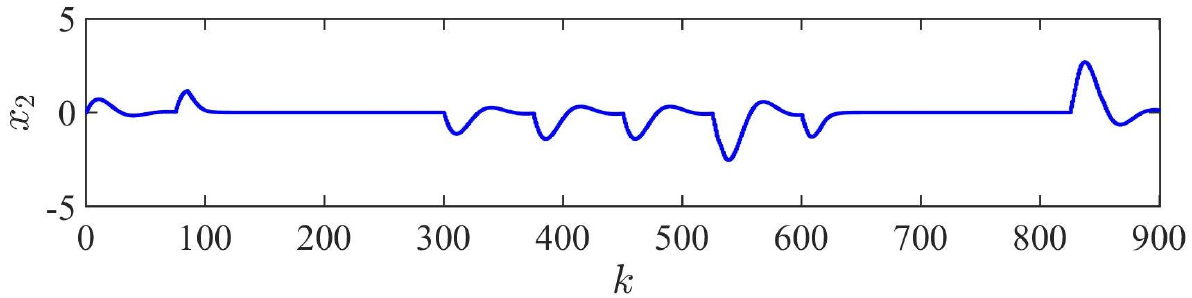}
	}
	\quad
	\subfigure{
		\includegraphics[width=0.5\textwidth]{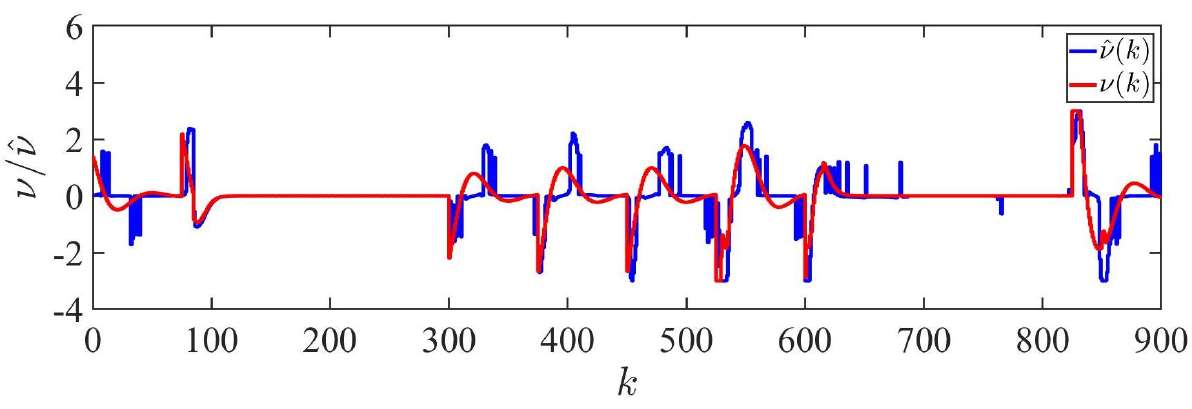}
	}\vspace{-0.3cm}
	\caption{Evolution of $x_2(k)$, input sequences $\nu(k)$ and $\hat{\nu}(k)$. }
	\label{fig:sim_uv_running}
\end{figure}
The explicit form of the obtained CBF $\mathcal{B}(x)$ and ACBF $\mathcal{B}_O(x,\hat{x})$ associated with the final $\mathcal{S}^{\text{exp}}$ and $\mathcal{S}_O^{\text{exp}}$, as well as the functions $k_a(x)$ and $\hat{k}_a(x,\hat{x})$ as in Theorem~\ref{thm:bilinear}, with $a=1$, are provided in Appendix~\ref{ap3}. 

To validate the obtained CBF and ACBF, we deploy the controller constructed based on the CBF and ACBF as described in Section~\ref{implementation}, considering the cost function as in~\eqref{cost}. 
We initialize the system at $x(0)=[3;0]$, and select $\hat{x}(0)=[2.77;0]$.
Then, we simulate the system for 900 time steps (90 sec.).
During the simulation, we change the set points as in~\eqref{cost} from time to time. 
The evolution of the set points and simulation results of the car are shown in Figure~\ref{fig:sim_runningexample}.
Between time step $k=100$ and $k=305$, the car enters the secrete state set $X^{\text{inf}}_s$. 
However, the intruder is not certain whether the car has actually entered the region $X^{\text{inf}}_s$ due to the existence of $\mathbf{x}_{\hat{x}(0),\hat\nu}$ that has not entered $X^{\text{inf}}_s$.
Between time step $k=600$ and $k=825$, a set point is given to drive the car away from the safety region.
Thanks to the synthesized CBF, the car does not go outside of the safety region so that the desired safety property is satisfied.
Meanwhile, as shown in Figure~\ref{fig:sim_uv_running}, the desired velocity and input constraint are also respected.

\subsection{Chaser Satellite}
\begin{figure}
	\begin{center}
		\includegraphics[width=0.4\textwidth]{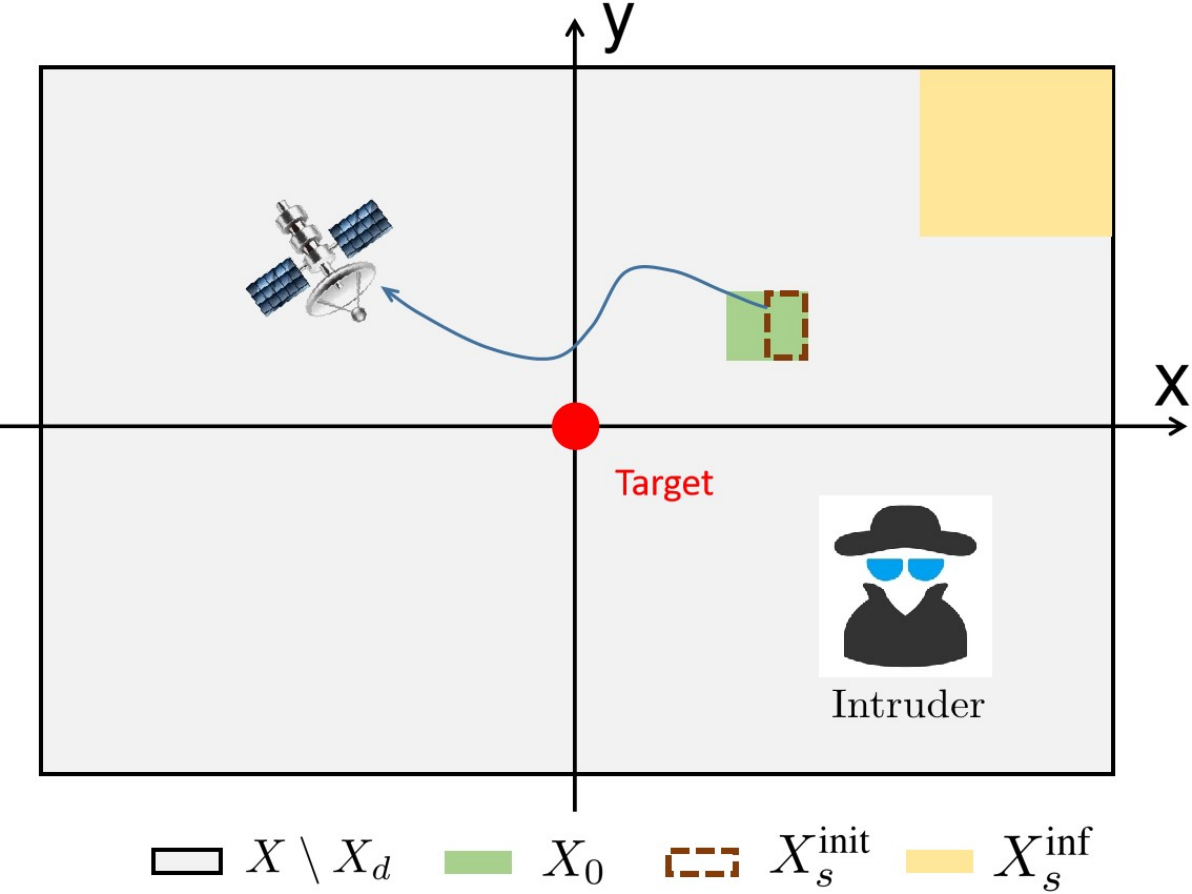}    
		\caption{A chaser satellite being observed by an intruder.} 
		\label{fig:satellite}
	\end{center}
\end{figure}

In this case study, we focus on controlling a chaser satellite, as shown in Figure~\ref{fig:satellite}.
The chaser satellite is moving around a target, while the position of the satellite is observed by a malicious intruder.
It is undesired to be revealed to the intruder whether or not
\begin{itemize}
	\item the satellite started from the region $X^{\text{init}}_s$;
	\item the satellite has ever entered the region $X^{\text{inf}}_s$.
\end{itemize}
Additionally, the desired safety region is characterized by 1) a neighborhood region around the target in which the chaser satellite must stay in; 2) velocity restriction within this region.
The motion of the chaser satellite can be modeled as follows, which is borrowed from~\cite{Jewison2016spacecraft}:
\begin{align*}
	x(k+1) &= Ax(k) + B\nu(k),\\
	y(k) &= Cx(k)
\end{align*}
with 
\begin{align*} 
	A := \begin{bmatrix}1& 0& 1.1460&  0.7038\\0.8156& 1& -0.0064& 4.3382\\
		0& 0& 0.8660& 1.1460\\0& 0& -0.2182& 0.8660	\end{bmatrix},
	B := \begin{bmatrix}0& 0.3031\\0.006& 0.1286&\\0& 0.0014\\0& 0.0023	\end{bmatrix} ,
	C := \begin{bmatrix} 1 & 0& 0& 0\\0& 0& 1& 0\end{bmatrix},
\end{align*} 
in which $x \!=\! [x_1;x_2;x_3;x_4]$ is the state of the system, with $x_1$ and $x_2$ (resp. $x_3$ and $x_4$) being the relative position (m) and velocity (m/s) between the satellite and the target on $\mathsf{x}$-axis (resp. $\mathsf{y}$-axis), respectively;
$u=[u_1;u_2]$ denotes the control input of the system, in which $u_1,u_2\in [-10,10]$ are the thrust force of the satellite on the $\mathsf{x}$-axis and $\mathsf{y}$-axis, respectively,
and $y$ is the output of the system observed by a malicious intruder, with observation precision $\delta =2$.
Here, we are interested in 
state set $X := [-45, 45] \times [-10,10]\times [-20,20]\times [-10,10]$,
initial set $X_0 := [3.6, 4.3] \times \{0\} \times [3.7, 4.5] \times \{0\}$,
secret sets $X^{\text{init}}_s:= [4, 4.3] \times \{0\} \times [3.7, 4.5] \times \{0\}$ and $X^{\text{inf}}_s:= [29, 35] \times [-5, 5]\times [10,12]\times [-5, 5]$,
and unsafe set $X_d:=X\setminus \big( [-35, 35] \times [-5,5]\times [-12,12]\times [-5,5] \big)$.
Considering $\epsilon = 0.01$, we construct the sets $\mathcal{R}^{\text{init}}_0$ as in~\eqref{set:init}, $\mathcal{R}^{\text{init}}_d$ as in~\eqref{set:iunsafe}, 
$\mathcal{R}^{\text{inf}}_0$ as in~\eqref{set:init_inf}, and $\mathcal{R}^{\text{inf}}_d$ as in~\eqref{set:isecure_inf}. 
Accordingly, as for the desired $\delta$-approximate initial-state opacity, we select $\mathcal{R}_0$ satisfying~\eqref{foraexi1}, denoted by $\mathcal{R}'_0$, as:
\begin{align}
	&\{x \!\in\! X^{\text{init}}_s, \hat{x}\in [-3.99, 0)\times \{0\}\times [3.99, 4.01)\times\{0\} \mid ||h(x) \!- \!h(\hat x)|| \!\leq\! \delta - \epsilon\}.\label{R0_init_sat}
\end{align}
As for the $\delta$-approximate infinite-step opacity, we choose $\mathcal{R}_0$ satisfying~\eqref{foraexi2}, denoted by $\mathcal{R}''_0$, as:
\begin{align}
	\!\!\!\!\!\!\{(x,\hat{x})\!\in\! (X_0\!\setminus\! X^{\text{init}}_s) \!\times\! (X_0\!\setminus\! X^{\text{init}}_s) \!\mid\!  ||h(x) \!- \!h(\hat x)|| \!\leq\! \delta \!-\! \epsilon\}.\!\!\!\!\label{R0_inf_sat}
\end{align}
To compute the sets $\mathcal{S}^{\text{init}}$ and $\mathcal{S}_O^{\text{init}}$, we choose 
\begin{itemize}
	\item the set $\bar{X}$ as in~\eqref{X_sub}, with 
	$a_1:=[0.0289;0;0;0]^\top$, $a_2:=[0;0.2020;0;0]^\top$, $a_3:=[0;0;0.0842;0]^\top$, $a_4:=[0;0;0;0.2020]^\top$, 
	$a_5:=-a_1$, $a_6:=-a_2$, $a_7:=-a_3$, and $a_8:=-a_4$;
	\item the set $\bar{\mathcal{R}}$ as in~\eqref{R_sub}, with 
	$b_1 := [0.7107; 0;0;0;  -0.7107; 0; 0; 0]^\top$,
	$b_2 := [0; 0; 0.7107; 0;0;0;-0.7107;   0]^\top$,
	$b_3 := [0; 0; 0; 0;0;0;0.1;0]^\top$,
	$b_4 := -b_1$, and $b_5 := -b_2$;
	\item candidates of $\bar{K}(x)$ and $\bar{K}_o(x,\hat{x})$ as in Theorem~\ref{opt_nonlinear_init}, with deg$(\bar{K}(x))=0$ and deg$(\bar{K}_o(x,\hat{x}))=0$.
\end{itemize}
Accordingly, we compute $\mathcal{S}^{\text{init}}$ and $\mathcal{S}_O^{\text{init}}$ by leveraging Theorem~\ref{opt_nonlinear_init} and Algorithm~\ref{alg:initial}, and we obtain
\begin{align*}
	\bar{K} = \begin{bmatrix} -1.259&   -0.844&   -0.363&   -7.442\\
		-5.242&   -3.2554 &  -1.417&  -31.044\end{bmatrix}\!,
	Q = \begin{bmatrix} 874.279 &-29.878& -8.132 &-144.055\\
		-29.878& 23.631&-42.309&4.542\\
		-8.132&-42.309&127.878&-0.025\\
		-144.055&4.542&-0.025&24.213
	\end{bmatrix}\!,
\end{align*}
with $c_1=c_2= 1$. Matrices $Q_o$ and $\bar{K}_o$ are provided Appendix~\ref{ap3}.

\begin{figure}
	\begin{center}
		\includegraphics[width=0.4\textwidth]{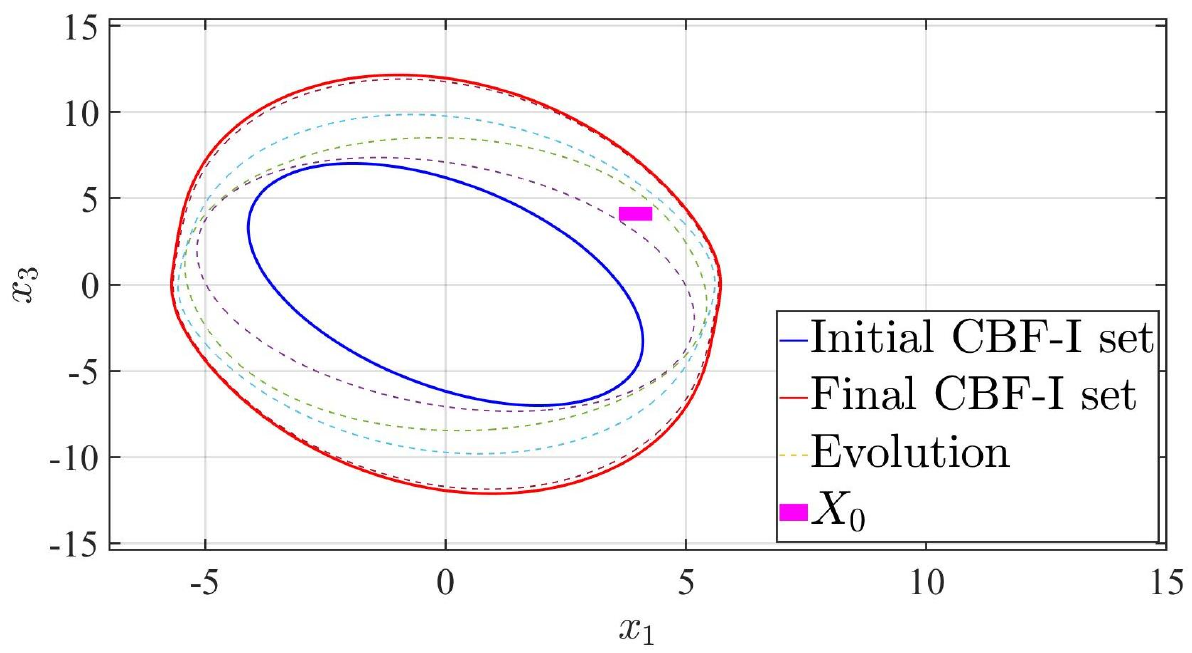}    
		\caption{Evolution of the cross section of the CBF-I set with respect to the number of iterations for the chaser satellite, with $x_2=x_4=0$.} 
		\label{fig:CBF_sate}
	\end{center}
\end{figure}

Having the initial CBF-I set $\mathcal{S}^{\text{init}}$ and the initial ACBF-I set $\mathcal{S}_O^{\text{init}}$ obtained by leveraging Theorem~\ref{opt_nonlinear_init}, 
we then deploy the iterative scheme proposed in Section~\ref{sec:iterative} to compute the expanded CBF-I set $\mathcal{S}^{\text{exp}}$ and the expanded ACBF-I set $\mathcal{S}_O^{\text{exp}}$.
Here, we consider candidates of $\mathcal{B}(x)$, $\mathcal{B}_O(x,\hat{x})$, $k_a(x)$, and $\hat{k}_a(x,\hat{x})$, in which $a\in\{1,2\}$, with deg$(\mathcal{B}(x))=4$, deg$(\mathcal{B}_O(x,\hat{x}))=2$, deg$(k_a(x))=2$, and deg$(\hat{k}_a(x,\hat{x}))=2$.
The computation ends in 5 iterations.
Here, we depict the evolution of the sets $\mathcal{S}^{\text{exp}}$ with respect to the number of iterations in Figure~\ref{fig:CBF_sate}.
\begin{figure}[t!]
	\centering
	\subfigure{
		\includegraphics[width=0.5\textwidth]{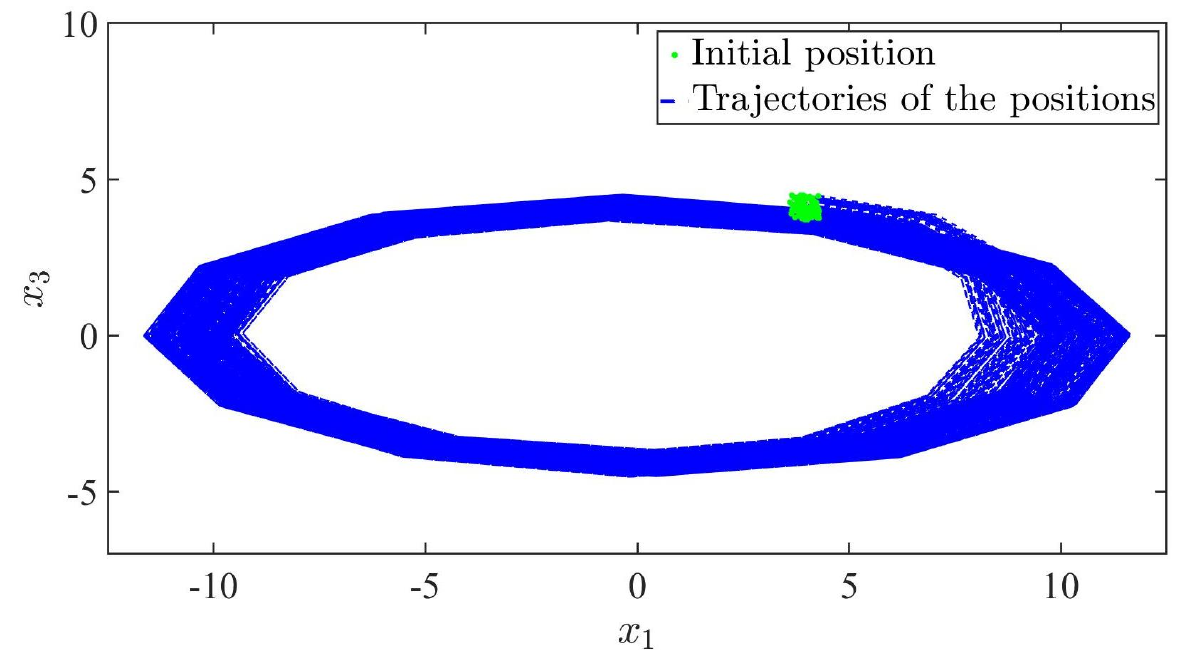}
	}
	\quad
	\subfigure{
		\includegraphics[width=0.5\textwidth]{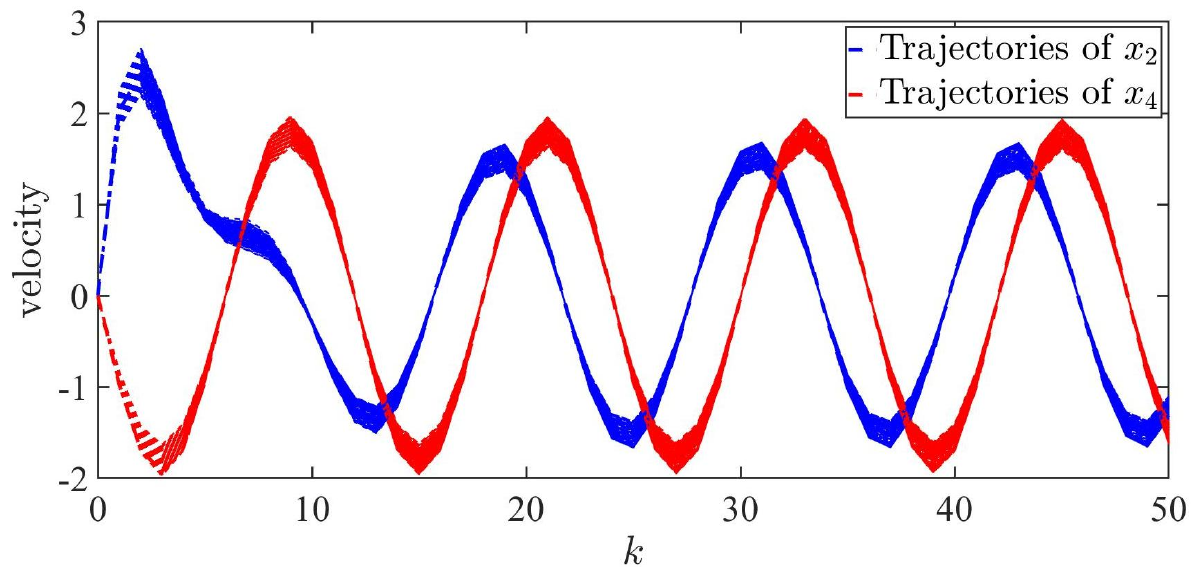}
	}
	\quad
	\subfigure{
		\includegraphics[width=0.5\textwidth]{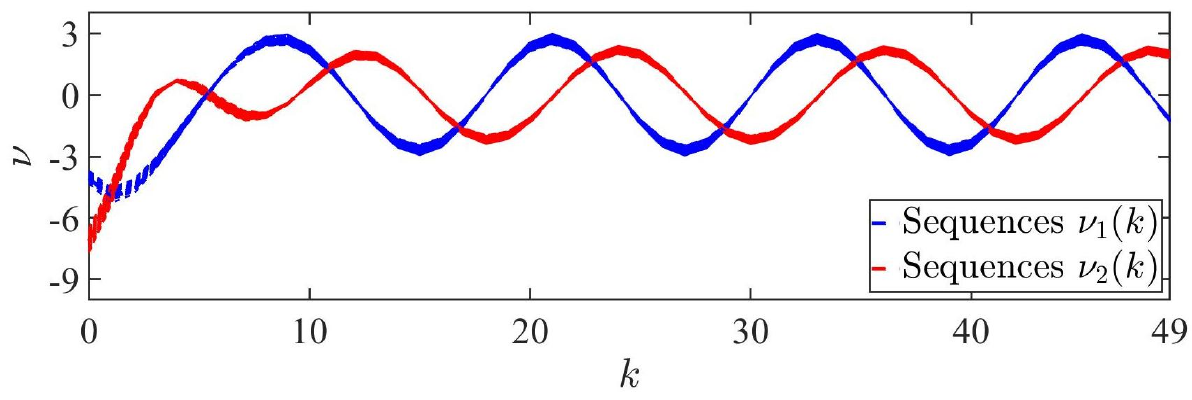}
	}
	\caption{Trajectories of positions, velocities, and control inputs of the chaser satellite in the simulations, indicating that the desired safety properties and input constraints are respected. }
	\label{fig:sim_satellite}
\end{figure}
Accordingly, the CBF $\mathcal{B}(x)$ and the ACBF $\mathcal{B}_O(x,\hat{x})$ associated with the final $\mathcal{S}^{\text{exp}}$ and $\mathcal{S}_O^{\text{exp}}$, as well as the functions $k_a(x)$ and $\hat{k}_a(x,\hat{x})$, as in Theorem~\ref{thm:bilinear},  with $a\in\{1,2\}$, are provided in Appendix~\ref{ap3}.

\begin{figure}
	\begin{center}
		\subfigure{
			\includegraphics[width=0.45\textwidth]{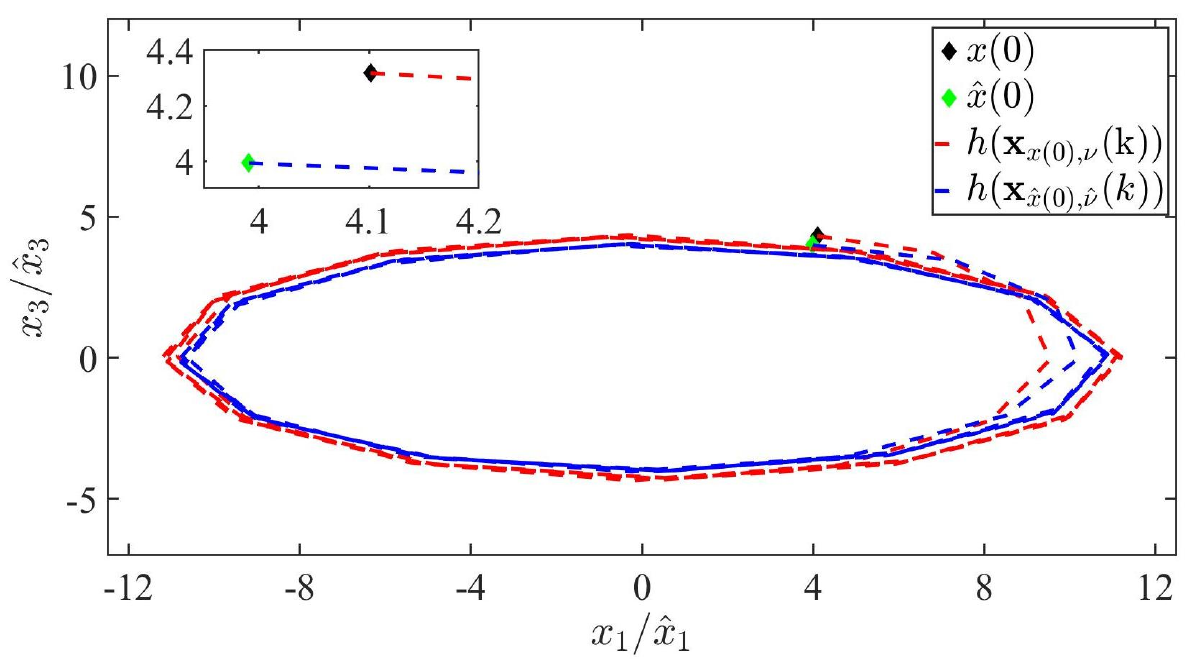}
		}
		\quad
		\subfigure{
			\includegraphics[width=0.45\textwidth]{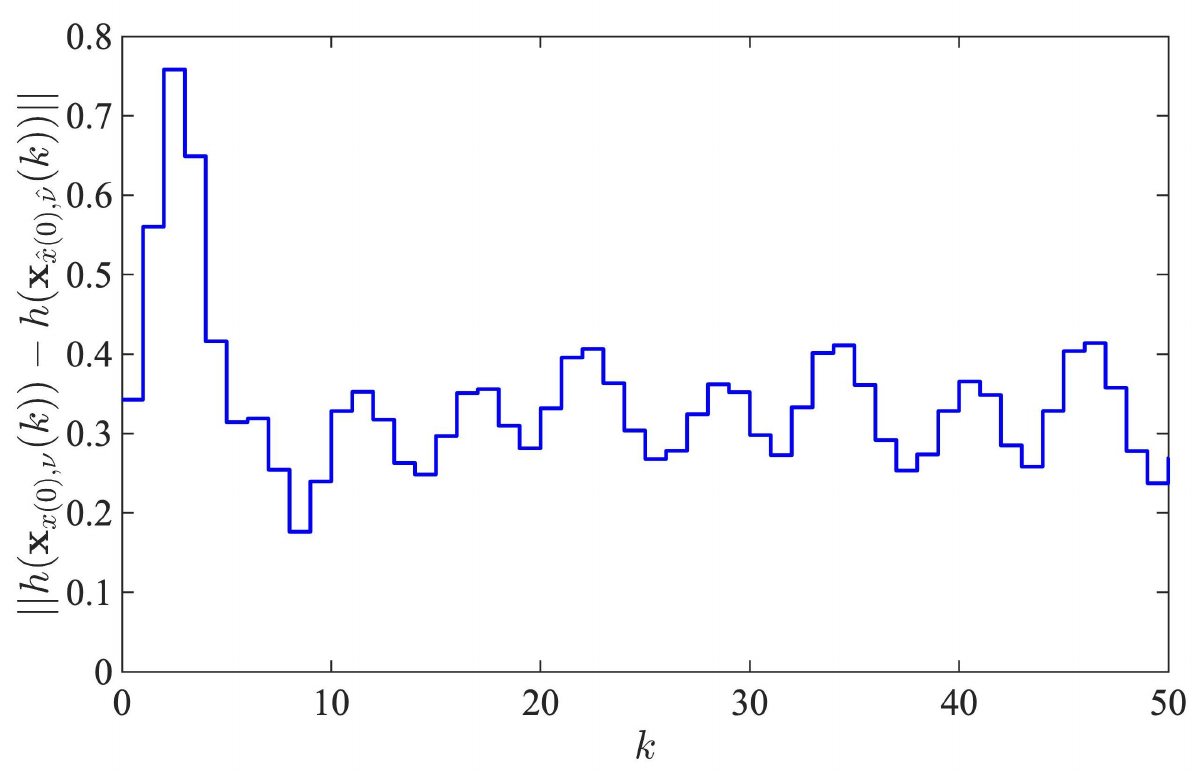}
		}
		\caption{
			A state run $\mathbf{x}_{x(0),\nu}$ of the system with its equivalent ($\delta$-close output) trajectory $\mathbf{x}_{\hat{x}(0),\hat{\nu}}$, in which $(x(0),\hat{x}(0))\in\mathcal{R}'_0 \cup \mathcal{R}''_0$, where $\mathcal{R}'_0$ and $\mathcal{R}''_0$ are defined as in~\eqref{R0_init_sat} and~\eqref{R0_inf_sat}, respectively. } 
		\label{fig:example_opacity}
	\end{center}
\end{figure}

\begin{figure}
	\begin{center}
		\includegraphics[width=0.45\textwidth]{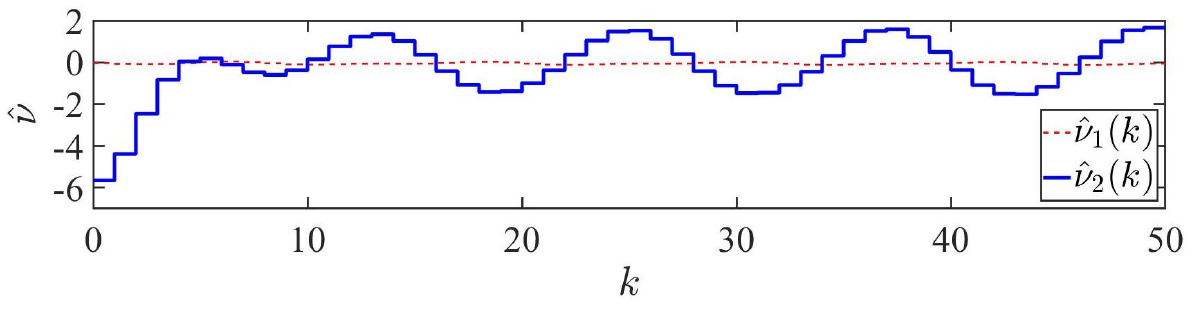}    
		\caption{Trajectory of $\hat \nu$ associated with $\mathbf{x}_{\hat{x}(0),\hat \nu}$ in Figure~\ref{fig:example_opacity}.}
		\label{fig:example_opacity_input}
	\end{center}
\end{figure}

To validate the obtained CBF and ACBF, we randomly select 100 initial states $x(0)$ from the initial state set $X_0$ and simulated the system for 50 time steps.
For those $x(0)\in X_0\cap X^{\text{init}}_s$, we randomly select $\hat{x}(0)$ such that $(x(0),\hat{x}(0))\in\mathcal{R}_0$, with $\mathcal{R}_0$ as in~\eqref{R0_init_sat}; for those $x(0)\in X_0\backslash X_s$, we set $\hat{x}(0)=x(0)$ considering the settings in~\eqref{R0_init_sat} and~\eqref{R0_inf_sat}.
Moreover, the secure-by-construction controller $k(x)$ associated with the CBF $\mathcal{B}(x)$ and the ACBF $\mathcal{B}_o(x,\hat{x})$ is deployed to control the chaser satellite in the simulation.
The state trajectories of the satellite and the corresponding input sequences $\nu(k)$, are depicted in Figure~\ref{fig:sim_satellite}, indicating that the desired safety properties and input constraints are respected.
Additionally, the desired $2$-approximate initial-state and infinite-step opacity are satisfied since for each collected $x(0)$ and its corresponding trajectory $\mathbf{x}_{x(0),\nu}$, there exists $\hat{x}(0)\in X_0\backslash X_s$ and trajectory $\mathbf{x}_{\hat{x}(0),\hat \nu}$, in which $\hat{\nu}(k)\in U$, such that $\Vert h(\mathbf{x}_{x(0),\nu}(k))-h(\mathbf{x}_{\hat{x}(0),\hat\nu}(k))\Vert \leq 2$ holds for all $k\in[0,50]$.
Additionally, all the trajectory pairs $(\mathbf{x}_{x(0),\nu},\mathbf{x}_{\hat{x}(0),\hat \nu})$ never reach $\mathcal{R}^{\text{inf}}_d$.
Here, we demonstrate in Figure~\ref{fig:example_opacity} an example of such a pair of trajectories.
Additionally, we also depict the sequence $\hat \nu$ associated with $\mathbf{x}_{\hat{x}(0),\hat \nu}$ in Figure~\ref{fig:example_opacity_input}.

\section{Conclusion}\label{sec5}
In this paper, we proposed a notion of (augmented) control barrier functions to construct secure-by-construction controllers enforcing safety and security properties over control systems \emph{simultaneously}. 
From safety perspective, we considered invariance properties, for which the system is expected to stay within a pre-specified safety set.
From security perspective, we focused on $\delta$-approximate initial-state and infinite-step opacity, which require that any intruder should never be certain whether the system originated from (resp. has visited) a designated secret state set based on their imperfect observations of the system's behavior.
Accordingly, we proposed conditions under which the (augmented) control barrier functions and their associated secure-by-construction controllers can be synthesized.
Given valid (augmented) control barrier functions, we introduced how to incorporate user-defined cost functions when designing the secure-by-construction controllers.
Additionally, we proposed an iterative sum-of-square programming to systematically compute (augmented) control barrier functions over polynomial control-affine systems.
In the future, we are interested to extend the current results by considering more general safety properties (e.g., those expressed as linear temporal logic formulae~\cite{Pnueli1977Temporal}, instead of simple invariance properties), and by exploring other notions of security properties \cite{dibaji2019systems} for complex CPS.

\bibliography{refr} 
\bibliographystyle{IEEEtran}

\appendix
\renewcommand{\theequation}{A.\arabic{equation}}
\renewcommand{\thetheorem}{A.\arabic{theorem}}

\section{Proof of Statements}

\subsection{Proof of the Results in Section~\ref{sec3}} \label{proof}   
\begin{figure*}
	\rule[0pt]{\textwidth}{0.05em}
	\begin{align}
		&\ \forall \mathsf{a}\in[1, n_{\mathsf{a}}], - \mathcal{B}(x)-\sum_{k=1}^{k_0(\mathsf{a})} s_{d,k}(\mathsf{a})\alpha_{\mathsf{a},k}(x)\in \mathcal{P}_S(x), \text{ with } s_{d,k}(\mathsf{a})\in \mathcal{P}_S(x), \forall k\in[1,k_0(\mathsf{a})];\label{cd4}\\
		&\ \forall \mathsf{c}\in[1, n_{\mathsf{c}}], s_{a,0}(\mathsf{c})\mathcal{B}(x)-\sum_{k=1}^{k_d(\mathsf{c})} s_{a,k}(\mathsf{c})\gamma_{\mathsf{c},k}(x)\in \mathcal{P}_S(x), \text{ with } s_{a,k}(\mathsf{c})\in \mathcal{P}_S(x), \forall k\in[1,k_d(\mathsf{c})];\label{cd1}\\
		&\ \forall \mathsf{d}\in[1, n_{\mathsf{d}}],  - \mathcal{B}_O(x,\hat{x})-\sum_{r=1}^{r(\mathsf{d})} \bar{s}_{d,r}(\mathsf{d})\kappa_{\mathsf{d},r}(x,\hat{x})\in \mathcal{P}_S(x,\hat{x}), \text{ with } \bar{s}_{d,r}(\mathsf{d})\in \mathcal{P}_S(x,\hat{x}), \forall r\in[1,r(\mathsf{d})];\label{cdd4}\\
		&\forall \mathsf{e}\in[1, n_{\mathsf{e}}], \bar{s}_{a,0}(\mathsf{e})\mathcal{B}_O(x,\hat{x})- \sum_{r=1}^{\mathsf{r}(\mathsf{e})} \bar{s}_{a,r}(\mathsf{e})\lambda_{\mathsf{e},r}(x,\hat{x})\in \mathcal{P}_S(x,\hat{x}), \text{ with } \bar{s}_{a,k}(\mathsf{e})\in \mathcal{P}_S(x,\hat{x}), \forall r\in[1,r(\mathsf{e})];\label{cdd1}\\
		&s_{b,1}\mathcal{B}(x)-s_{b,2}\mathcal{B}(f(x,u))-\sum_{a=1}^{m}p_b(a)(u_a - k_a(x))\!\in\! \mathcal{P}_S(x,u),\nonumber\\
		&\quad \quad \quad \quad \quad  \quad \quad \quad \quad  \quad \quad \quad \quad  \quad\text{with } u\!=\![u_1;\ldots;u_m], s_{b,1},s_{b,2}\!\in\! \mathcal{P}_S(x,u); p_b(a)\!\in\!\mathcal{P}(x,u), \forall a \!\in\! [1,m];\label{cd2}\\
		&\bar{s}_{b,1}\mathcal{B}_O(x,\hat{x})\!-\!\bar{s}_{b,2}\mathcal{B}_O(f(x,u),f(\hat{x},\hat{u}))\!-\!\sum_{a=1}^{m}\bar{p}_b(a)(u_a \!- \!k_a(x))-\sum_{a=1}^{m}\hat{p}_b(a)(\hat{u}_a \!-\! \hat{k}_a(x,\hat{x}))\!\in\! \mathcal{P}_S(x,\hat{x},u,\hat{u}),\nonumber\\
		&\quad  \text{with } u\!=\![u_1;\ldots;u_m], \hat{u}\!=\![\hat{u}_1;\ldots;\hat{u}_m]; \bar{s}_{b,1},\bar{s}_{b,2}\!\in\! \mathcal{P}_S(x,\hat{x},u,\hat{u}); \bar{p}_b(a),\hat{p}_b(a)\!\in\!\mathcal{P}(x,\hat{x},u,\hat{u}), \forall a \!\in\! [1,m];\label{cdd2}\\		
		&\forall j \in [1,\mathsf{j}], s_{c,1}(j)\mathcal{B}(x)-s_{c,2}(j)\rho_{j}(u)+s_{c,3}(j)\rho_{j}(u)\mathcal{B}(x)-\sum_{a=1}^{m}p_c(a,j)(u_a - k_a(x))-\rho_{j}(u)^{2k}\in\mathcal{P}_S (x,u),\nonumber\\
		&\quad \quad \quad \quad   \text{with }u=[u_1;\ldots;u_m], k\in\mathbb{N}, s_{c,1}(j),s_{c,2}(j),s_{c,3}(j)\in \mathcal{P}_S(x,u), p_c(a,j)\in\mathcal{P}(x,u), \forall a \in [1,m];\label{cd3}\\
		&\forall j \!\in\! [1,\mathsf{j}], \bar{s}_{c,1}(j)\mathcal{B}_O(x,\hat{x})\!-\!\bar{s}_{c,2}(j)\rho_{j}(\hat{u})\!+\!\bar{s}_{c,3}(j)\rho_{j}(\hat{u})\mathcal{B}_O(x,\hat{x})\!-\!\sum_{a=1}^{m}\bar{p}_c(a,j)(\hat{u}_a \!-\! \hat{k}_a(x,\hat{x}))\!-\!\rho_{j}(\hat{u})^{2\bar{k}}\in\!\mathcal{P}_S(x,\hat{x},\hat{u}), \nonumber\\
		&~~~~~~~~text{with }\hat{u}=[\hat{u}_1;\ldots;\hat{u}_m],\bar{k}\in\mathbb{N}, \bar{s}_{c,1}(j),\bar{s}_{c,2}(j),\bar{s}_{c,3}(j)\in \mathcal{P}_S(x,\hat{x},\hat{u}), \bar{p}_c(a,j)\in\mathcal{P}(x,\hat{x},\hat{u}), \forall a \in [1,m];\label{cdd3}
	\end{align}
	\rule[0pt]{\textwidth}{0.05em}
	\begin{align}
		&\{x,\in\mathbb{R}^n, u\in\mathbb{R}^m|\mathcal{B}(x)\leq 0 \}\cap \bigcap_{a=1}^{\mathsf{m}}
		\left\{x\in\mathbb{R}^n, u\in\mathbb{R}^m | u_a = k_a(x), \text{ with } u:=[u_1;\ldots,u_m]
		\right\}\nonumber\\ 
		& \quad  \quad\quad\quad \quad \quad \quad \quad \quad\quad\quad \quad \quad \quad \quad \quad \quad \quad \quad\quad\quad\quad \quad \quad \quad\quad\quad\quad\subseteq 
		\{x\in\mathbb{R}^n,u\in\mathbb{R}^m| \mathcal{B}(f(x,u))\leq 0\};\label{cdd01}\\
		&\{x, \hat{x}\in\mathbb{R}^n, u,\hat{u}\in\mathbb{R}^m|\mathcal{B}_O(x,\hat{x})\leq 0 \}\cap \bigcap_{a=1}^{\mathsf{m}}
		\left\{\begin{aligned}x,\hat{x}\in\mathbb{R}^n,\\ u,\hat{u}\in\mathbb{R}^m	\end{aligned} \Bigg|
		\begin{aligned}
			&u_a = k_a(x), \text{ with } u:=[u_1;\ldots,u_m];\\&\hat{u}_a = \hat{k}_a(x,\hat{x}), \text{ with } \hat{u}:=[\hat{u}_1;\ldots,\hat{u}_m].
		\end{aligned}\right\}\nonumber\\ 
		& \quad  \quad\quad\quad \quad \quad \quad \quad \quad\quad\quad \quad \quad \quad \quad \quad \quad \quad \quad \quad\quad\subseteq 
		\{x, \hat{x}\in\mathbb{R}^n,u,\hat{u}\in\mathbb{R}^m| \mathcal{B}_O\big(f(x,u),f(\hat{x},\hat{u})\big)\leq 0\};\label{cd01}\\
		&\{x\!\in\! \mathbb{R}^n, u \!\in\! \mathbb{R}^m | \mathcal{B}(x)\!\leq\! 0, u_a\! =\! k_a(x),\forall a\!\in\![1,m], \text{ with } u\!:=\![u_1;\ldots,u_m]\} \!\subseteq\! \bigcap_{j=1}^{\mathsf{j}}\{x\in \mathbb{R}^n,\, u\!\in\! \mathbb{R}^m\big|\rho_{j}(u)\!\leq\! 0 \};\label{cd02}\\
		&\left\{x,\hat{x}\!\in\! \mathbb{R}^n, \hat{u} \!\in\! \mathbb{R}^m \Bigg| 
		\begin{aligned}	&\mathcal{B}_O(x,\hat{x})\!\leq\! 0;\hat{u}_a\! =\! \hat{k}_a(x,\hat{x}),\\	&\forall a\!\in\![1,m], \text{ with } \hat{u}\!:=\![\hat{u}_1;\ldots,\hat{u}_m]
		\end{aligned}\right\} \!\subseteq\! \bigcap_{j=1}^{\mathsf{j}}\{x,\hat{x}\in \mathbb{R}^n,\, \hat{u}\!\in\! \mathbb{R}^m\big|\rho_{j}(\hat{u})\!\leq\! 0 \};\label{cd03}\\
		&\left\{ x\in\mathbb{R}^n,u\in\mathbb{R}^m  |\mathcal{B}(x) \leq 0,\mathcal{B}\big(f(x,u)\big)\geq 0;	u_a = k_a(x), a\in[1,m], \text{ with } u:=[u_1;\ldots,u_m], 
		\right\}=\emptyset;  \label{cddh01}\\
		&\left\{ \begin{aligned}x, \hat{x}\in\mathbb{R}^n,\\u,\hat{u}\in\mathbb{R}^m \end{aligned} \Bigg|
		\begin{aligned}
			&\mathcal{B}_O(x,\hat{x}) \leq 0,\mathcal{B}_O\big(f(x,u),f(\hat{x},\hat{u})\big)\geq 0;\\
			&u_a = k_a(x), \hat{u}_a = \hat{k}_a(x,\hat{x}), a\in[1,m], \text{ with } u:=[u_1;\ldots,u_m], \hat{u}:=[\hat{u}_1;\ldots,\hat{u}_m]
		\end{aligned}\right\}=\emptyset;  \label{cdh01}\\
		& \forall j\in [1,\mathsf{j}], \{x\!\in\! \mathbb{R}^n, u \!\in\! \mathbb{R}^m | \mathcal{B}(x)\leq 0; u_a\! =\! k_a(x),\forall a\!\in\![1,m], \text{ with } u\!:=\![u_1;\ldots,u_m];\rho_{j}(u)\!\geq\! 0 , \rho_{j}(u)\!\neq\! 0\}=\emptyset;\label{cdh02}\\
		& \forall j\in [1,\mathsf{j}],\left\{x,\hat{x}\!\in\! \mathbb{R}^n, \hat{u} \!\in\! \mathbb{R}^m \Bigg| 
		\begin{aligned}	&  \mathcal{B}_O(x,\hat{x})\!\leq\! 0,\rho_{j}(\hat{u})\!\geq\! 0 , \rho_{j}(\hat{u})\!\neq\! 0\\	&\hat{u}_a\! =\! \hat{k}_a(x,\hat{x}),\forall a\!\in\![1,m], \text{ with } \hat{u}\!:=\![\hat{u}_1;\ldots,\hat{u}_m]
		\end{aligned}\right\} =\emptyset.\label{cdh03}
	\end{align}
	\rule[0pt]{\textwidth}{0.05em}
\end{figure*}

{\bf Proof of Theorem~\ref{BC_1}.}
We first show that $C(x)$ enforces the desired safety property.
Consider the functions $\mathcal{B}$ and $C$ such that conditions (\textbf{Cond.1}) and (\textbf{Cond.5}) in Definition~\ref{cond} hold, then, starting from any initial state $x_0\in X_0$, one has $\mathbf{x}_{x_0,\nu}(k)\in \mathcal{S}$, for all $k\in\mathbb{N}$, with $\nu$ being generated by $C$.
On the other hand, (\textbf{Cond.2}) ensures that $\mathcal{S}\cap X_d = \emptyset$,  indicating that from any initial state $x_0\in X_0$, one has $\mathbf{x}_{x_0,\nu}(k)\notin X_d $, for all $k\in\mathbb{N}$.
Hence, $C(x)$ enforces the desired safety properties in Problem~\ref{prob}.

Next, we show that with sets $\mathcal{R}^{\text{inf}}_{0}$ and $\mathcal{R}^{\text{inf}}_{d}$ defined as in \eqref{set:init_inf} and \eqref{set:isecure_inf}, respectively, $C(x)$ also enforces approximate infinite-step opacity of $\Sigma$ as in Definition~\ref{def:opa}.
Consider any arbitrary initial state $x_0 \in X_0$ and an arbitrary input sequence $\nu$ generated by the controller $C$, and the corresponding state trajectory  $\mathbf{x}_{x_0,\nu}\!=\!(x_0,\dots, x_n)$ such that $x_k \in X^{\text{inf}}_s$ for some $k \in [0,n]$. 
We consider the following two cases:
\begin{enumerate}
	\item If $k=0$, then we have $x_0 \in X^{\text{inf}}_s$. By the assumption as in \eqref{initassum} that $\{x \!\in\! X_0 | \Vert h(x) - h(x_0)\Vert \!\leq\! \delta \} \nsubseteq X^{\text{inf}}_s$ for all $x_0 \!\in\! X_0 \cap X^{\text{inf}}_s$, we get that there exists $\hat x_0 \in X \setminus X^{\text{inf}}_s$ such that $\Vert h(x) - h(x_0)\Vert \!\leq\! \delta$. 
	This implies that $\mathcal{R}^{\text{inf}}_{0}$ in~\eqref{set:init_inf} is not empty.
	Note that set $\mathcal{R}_0 \neq \emptyset$ satisfies
	$\overline{\textbf{Proj}}(\mathcal{R}^{\text{inf}}_0) \subseteq \overline{\textbf{Proj}}(R_0)$ as in \eqref{foraexi2}, we get that for any  $x_0 \!\in\! X_0 \cap X^{\text{inf}}_s$,  there exists $\hat x_0 \in \mathbb{R}^n$ such that $(x_0, \hat x_0) \in \mathcal{R}_0$.  Moreover, since $\underline{\textbf{Proj}}(R_0) \subseteq \underline{\textbf{Proj}}(\mathcal{R}^{\text{inf}}_0)$ holds as in \eqref{foraexi2}, we further have $\hat x_0 \in X \setminus X^{\text{inf}}_s$. Therefore, by (\textbf{Cond.3}), we get that  
	\begin{equation}
 	\medmuskip=1.3mu
	\thinmuskip=1.3mu
	\thickmuskip=1.3mu
		\!\!\!\!\!\forall x_0\!\in\! X_0 \!\cap\! X^{\text{inf}}_s, \!\exists\hat{x}_0 \!\in\!  X_0 \!\!\setminus\!\! X^{\text{inf}}_s, \text{s.t. } \Vert h(x_0)\! -\! h(\hat{x}_0)\Vert \!\leq\! \delta,\!\! \label{help123}
	\end{equation}
	with $(x_0,\hat{x}_0)\in \mathcal{S}_O$.
	Furthermore, since (\textbf{Cond.6}) holds, for $(x_0,\hat{x}_0)\in \mathcal{S}_O$, by deploying $u_0\in C(x_0)$, there exists $\hat{u}_0\in U$ such that $(x_1,\hat{x}_1)\in \mathcal{S}_O$, with $x_1 = f(x_0,u_0)$ and $\hat{x}_1=f(\hat{x}_0,\hat{u}_0)$.
	By induction, (\textbf{Cond.6}) implies that if one has 
	$(x_0,\hat{x}_0)\in \mathcal{S}_O$, there exists input runs $\nu:=(u_0,\ldots,u_k,\ldots)$ and $\hat{\nu}:=(\hat{u}_0,\ldots,\hat{u}_k,\ldots)$, such that one gets 
	$(\mathbf{x}_{x_0,\nu}(k),\mathbf{x}_{\hat x_0,\hat \nu}(k))\in \mathcal{S}_O$, for all $ k\in\mathbb{N}$, in which $\nu$ is generated by controller $C$.
	Moreover, (\textbf{Cond.4}) indicates that for all $(x,\hat{x})\in\mathcal{S}_O$, one gets $(x,\hat{x})\notin\mathcal{R}_d = \mathcal{R}^{\text{inf}}_d$, which implies that $\Vert h(x)-h(\hat{x}) \Vert\leq\delta$ by the structure of $\mathcal{R}^{\text{inf}}_d$ as in ~\eqref{set:isecure_inf}.  
	Therefore, one can conclude that for any $x_0 \!\in\!  X_0 \cap  X^{\text{inf}}_s$, for all $\mathbf{x}_{x_0,\nu}$ with $\nu$ being generated by $C(x)$, there exists $\mathbf{x}_{\hat x_0,\hat \nu}$ with $\hat{x}_0 \!\in\!  X_0 \setminus \!  X^{\text{inf}}_s$ such that one has $\Vert h(\mathbf{x}_{x_0,\nu}(k))-h(\mathbf{x}_{\hat x_0,\hat \nu}(k))\Vert \leq \delta$.
	\item If $k\geq 1$, then we have $x_0 \in X_0 \setminus X^{\text{inf}}_s$. 
	One can readily get that there exists $\hat x_0 \in X_0 \setminus X^{\text{inf}}_s$, with $\Vert h(x_0) - h(\hat x_0)\Vert \!\leq\! \delta$, such that $(x_0, \hat x_0) \in \mathcal{R}^{\text{inf}}_{0}$, with $\mathcal{R}^{\text{inf}}_{0}$ defined as in~\eqref{set:init_inf}.
	Moreover,~\eqref{foraexi2} implies that $(x_0, \hat x_0) \in \mathcal{R}_0$ so that one has $(x_0,\hat{x}_0)\in \mathcal{S}_O$ by leveraging (\textbf{Cond.3}).
	Then, since we have (\textbf{Cond.6}) holds, for the pair of states $(x_0,\hat{x}_0)\in \mathcal{S}_O$, by deploying $u_0\in C(x_0)$, there exists $\hat{u}_0\in U$ such that $(x_1,\hat{x}_1)\in \mathcal{S}_O$, with $x_1 = f(x_0,u_0)$ and $\hat{x}_1=f(\hat{x}_0,\hat{u}_0)$.
	By induction,  (\textbf{Cond.6}) implies that if one has  $(x_0,\hat{x}_0)\in \mathcal{S}_O$, there exists input runs $\nu:=(u_0,\ldots,u_k,\ldots)$ and $\hat{\nu}:=(\hat{u}_0,\ldots,\hat{u}_k,\ldots)$, such that one gets 
	$(\mathbf{x}_{x_0,\nu}(k),\mathbf{x}_{\hat x_0,\hat \nu}(k))\in \mathcal{S}_O$, for all $ k\in\mathbb{N}$, in which $\nu$ is generated by controller $C$.
	Moreover, (\textbf{Cond.4}) indicates that for all $(x,\hat{x})\in\mathcal{S}_O$, one gets $(x,\hat{x})\notin\mathcal{R}_d = \mathcal{R}^{\text{inf}}_d$, which implies that $\Vert h(x)-h(\hat{x}) \Vert\leq\delta$ by the structure of $\mathcal{R}^{\text{inf}}_d$ as in ~\eqref{set:isecure_inf}.  
	Note that for the secret state $x_k = \mathbf{x}_{x_0,\nu}(k)  \in X^{\text{inf}}_s$, by the structure of $\mathcal{R}^{\text{inf}}_d$ in ~\eqref{set:isecure_inf}, it follows that $\mathbf{x}_{\hat x_0,\hat \nu}(k) \in X \setminus X^{\text{inf}}_s$ with $\Vert h(\mathbf{x}_{x_0,\nu}(k))-h(\mathbf{x}_{\hat x_0,\hat \nu}(k)) \Vert\leq\delta$.
	Therefore, one can conclude that for any $x_0 \!\in\!  X_0 $, for all state trajectory $\mathbf{x}_{x_0,\nu}$ such that $x_k \in X^{\text{inf}}_s$, with $\nu$ being generated by $C(x)$, there exists $\mathbf{x}_{\hat x_0,\hat \nu}$ with $\hat{x}_0 \!\in\!  X_0$ and $\hat x_k \in X \setminus X^{\text{inf}}_s$ such that one has $\Vert h(\mathbf{x}_{x_0,\nu}(k))-h(\mathbf{x}_{\hat x_0,\hat \nu}(k))\Vert \leq \delta$ for all $k \in \mathbb{N}$. 
	This indicates that $C(x)$ enforces the approximate infinite-step opacity property as in Problem~\ref{prob} as well, which completes the proof. $\hfill\blacksquare$
\end{enumerate}

{\bf Proof of Corollary~\ref{inf_nocheck}.}
Note that~\eqref{maxmin2} holds indicates that~\eqref{help123} holds. 
The rest of the proof can be formulated similarly to that for Theorem~\ref{BC_1}. $\hfill\blacksquare$

{\bf Proof of Theorem~\ref{BC}.}
First note that by following the same proof of Theorem~\ref{BC_1}, a controller $C(x)$ satisfying (\textbf{Cond.1}), (\textbf{Cond.2}) and (\textbf{Cond.5}) as in Definition \ref{cond} enforces the safety property as in Problem \ref{prob}. 
Next, we proceed with showing that $C(x)$ also enforces the desired approximate initial-state opacity property as in Definition~\ref{def:opa}.
Consider any arbitrary secret initial state $x_0 \!\in\!  X_0 \cap  X^{\text{init}}_s$. 
By the assumption as in \eqref{initassum} that $\{x \!\in\! X_0 | \Vert h(x) - h(x_0)\Vert \!\leq\! \delta \} \nsubseteq X^{\text{init}}_s$ for all $x_0 \!\in\! X_0 \cap X^{\text{init}}_s$, we get that there exists $\hat{x}_0  \in X_0 \setminus X^{\text{init}}_s$ such that $\Vert h(\hat{x}_0) - h(x_0)\Vert \!\leq\! \delta$.
This indicates that the set $\mathcal{R}^{\text{init}}_0$ in~\eqref{set:init} is not empty.
Notice that since set $\mathcal{R}_0 \neq \emptyset$ satisfies $\overline{\textbf{Proj}}(\mathcal{R}^{\text{init}}_0) \subseteq \overline{\textbf{Proj}}(R_0)$  as in ~\eqref{foraexi1}, we get that for any $x_0\!\in\! X_0 \!\cap\! X^{\text{init}}_s$, there exists $\hat{x}_0 \in \mathbb{R}^n$ such that $(x_0, \hat{x}_0) \in \mathcal{R}_0$. Moreover, given that $\underline{\textbf{Proj}}(R_0) \subseteq \underline{\textbf{Proj}}(\mathcal{R}^{\text{init}}_0)$ as in ~\eqref{foraexi1}, we further have $\hat{x}_0 \in X_0 \setminus X^{\text{init}}_s$ holds. 
Therefore, by (\textbf{Cond.3}), we get that 
\begin{equation*}
	\!\!\!\!\!\forall x_0\!\in\! X_0 \!\cap\! X^{\text{init}}_s, \!\exists\hat{x}_0 \!\in\!  X_0 \!\!\setminus\!\! X^{\text{init}}_s, \text{s.t. } \Vert h(x_0)\! -\! h(\hat{x}_0)\Vert \!\leq\! \delta,\!\!
\end{equation*}
with  $(x_0,\hat{x}_0)\in \mathcal{S}_O$.
Moreover, since we further have (\textbf{Cond.6}) holds, for $(x_0,\hat{x}_0)\in \mathcal{S}_O$, by deploying $u_0\in C(x_0)$, there exists $\hat{u}_0\in U$ such that  $(x_1,\hat{x}_1)\in \mathcal{S}_O$, with $x_1 = f(x_0,u_0)$ and $\hat{x}_1=f(\hat{x}_0,\hat{u}_0)$.
By induction, (\textbf{Cond.6}) implies that if one has  $(x_0,\hat{x}_0)\in \mathcal{S}_O$, there exists input runs $\nu:=(u_0,\ldots,u_k,\ldots)$ and $\hat{\nu}:=(\hat{u}_0,\ldots,\hat{u}_k,\ldots)$, such that one gets  
$(\mathbf{x}_{x_0,\nu}(k),\mathbf{x}_{\hat x_0,\hat \nu}(k))\in \mathcal{S}_O$, for all $ k\in\mathbb{N}$, in which $\nu$ is generated by controller $C$.
On the other hand, (\textbf{Cond.4}) indicates that for all $(x,\hat{x})\in\mathcal{S}_O$, one gets $(x,\hat{x})\notin\mathcal{R}_d = \mathcal{R}^{\text{init}}_d$, which implies that $\Vert h(x)-h(\hat{x}) \Vert\leq\delta$ by the structure of $\mathcal{R}^{\text{init}}_d$ as in \eqref{set:iunsafe}.  
Therefore, one can conclude that for any $x_0 \!\in\!  X_0 \cap  X^{\text{init}}_s$, for all $\mathbf{x}_{x_0,\nu}$ with $\nu$ being generated by $C(x)$, there exists $\mathbf{x}_{\hat x_0,\hat \nu}$ with $\hat{x}_0 \!\in\!  X_0 \setminus \!  X^{\text{init}}_s$ such that one has $\Vert h(\mathbf{x}_{x_0,\nu}(k))-h(\mathbf{x}_{\hat x_0,\hat \nu}(k))\Vert \leq \delta$.
This indicates that $C(x)$ enforces the approximate initial-state opacity property as in Problem~\ref{prob} as well, which completes the proof. $\hfill\blacksquare$

{\bf Proof of Corollary~\ref{controller}}: 
{Here, we show Corollary~\ref{controller} by showing that for all $x_0\in X_0$, there exist $\hat{x}_0\in X_0$ and $\mathbf{x}_{\hat{x}_0,\nu}$, such that 
	\begin{align}
		&\mathbf{x}_{x_0,\nu}(k)\notin X_d,\forall k\in\mathbb{N};\label{imp:safety}\\
		&\Vert h(\mathbf{x}_{x_0,\nu}(k))-h(\mathbf{x}_{\hat x_0,\hat \nu}(k))\Vert \leq \delta,\forall k\in\mathbb{N};\label{imp:opacity}\\
		&x_0 \in X^{\text{init}}_s \implies \hat{x}_0 \notin X^{\text{init}}_s;\label{imp:initopa}\\
		&\mathbf{x}_{x_0,\nu}(k) \in X^{\text{inf}}_s \implies \hat{\mathbf{x}}_{x_0,\nu}(k) \notin X^{\text{inf}}_s ,\forall k\in\mathbb{N};\label{imp:infopa}
	\end{align}
	in which $\nu(k):=u$ and $\hat{\nu}(k):=\hat{u}$ are computed by solving the following optimization problem OP at each time step $k\in\mathbb{N}$.
	As a key insight, ~\eqref{imp:safety} corresponds to the desired safety property as in Problem~\ref{prob}, 
	~\eqref{imp:opacity} and~\eqref{imp:initopa} correspond to the desired opacity properties associated with $X^{\text{init}}_s$, 
	while~\eqref{imp:opacity} together with~\eqref{imp:infopa} indicate that the desired opacity properties associated with $X^{\text{inf}}_s$ are satisfied.}

Firstly, $\forall x_0\in X_0$, one has $x_0 \in \mathcal{S}$ considering (\textbf{Cond.1}).
Moreover, by applying input $\nu(k)$ satisfying $\mathcal{B}\big(f(\mathbf{x}_{x_0,\nu}(k),\nu(k))\big)\leq 0$ for all $k\in\mathbb{N}$  (existence of such $u(k)$ is guaranteed by (\textbf{Cond.5})), one has $\mathbf{x}_{x_0,\nu}(k) \in\mathcal{S}$ for all $k\in\mathbb{N}$ considering  and (\textbf{Cond.5}). 
Meanwhile, since one has $\mathcal{X}\cap X_d = \emptyset$ according to (\textbf{Cond.2}). 
Hence, one has~\eqref{imp:safety} holds. 
Next, we proceed with discussing~\eqref{imp:opacity}-\eqref{imp:infopa}:
\begin{itemize}
	\item If $x(0) = x_0\in X_0\cap X^{\text{inf}}_s$, one can select $\hat{x}(0)\in X_0\backslash X^{\text{inf}}_s$, such that $\Vert h(x(0))-h(\hat{x}(0))\Vert \leq \delta$ and $(x(0),\hat{x}(0))\in\mathcal{S}_O$ considering (\textbf{Cond.3}) and ~\eqref{foraexi2}.
	On one hand, by applying input $\nu(k)$ (and $\hat{\nu}(k)$) satisfying $\mathcal{B}\big(f(\mathbf{x}_{x_0,\nu}(k),\nu(k))\big)\leq 0$ and $\mathcal{B}_O\big(f(\mathbf{x}_{x_0,\nu}(k),\nu(k)),f(\hat{\mathbf{x}}_{x_0,\nu}(k),\hat{\nu}(k))\big)\leq 0$  (existence of such $\nu(k)$ and $\hat{\nu}(k)$ is guaranteed by (\textbf{Cond.5}) and (\textbf{Cond.6})), one has $(\mathbf{x}_{x_0,\nu}(k),\hat{\mathbf{x}}_{x_0,\nu}(k))\in\mathcal{S}_O$ holds for all $k\in\mathbb{N}$ according to (\textbf{Cond.5}) and (\textbf{Cond.6}).
	One the other hand, one has $\mathcal{S}_O \cap \mathcal{R}_d=\emptyset$ due to (\textbf{Cond.4}).
	Therefore, considering~\eqref{foraexi2}, one has~\eqref{imp:opacity} and~\eqref{imp:infopa} holds for all $k\in\mathbb{N}$.
	\item If $x(0) =x_0\in X_0\backslash X^{\text{inf}}_s$, there exists $\hat{x}(0) \in X_0$, with $\Vert h(x(0))-h(\hat{x}(0))\Vert \leq \delta$ and $(x(0),\hat{x}(0))\in\mathcal{S}_O$ considering (\textbf{Cond.3}) and ~\eqref{foraexi2}. 
	Then, one can show~\eqref{imp:opacity} and~\eqref{imp:infopa} also holds for all $k\in\mathbb{N}$ in a similar way as the case in which $x(0) \in X_0\cap X^{\text{inf}}_s$.
	\item If $x(0) = x_0\in X_0\cap X^{\text{init}}_s$, one can select $\hat{x}(0)\in X_0\backslash X^{\text{init}}_s$, such that $\Vert h(x(0))-h(\hat{x}(0))\Vert \leq \delta$ and $(x(0),\hat{x}(0))\in\mathcal{S}_O$ considering (\textbf{Cond.3}) and ~\eqref{foraexi1}.
	Then, one can show~\eqref{imp:opacity} and~\eqref{imp:initopa} also holds for all $k\in\mathbb{N}$ in a similar way as the case in which $x(0) \in X_0\cap X^{\text{inf}}_s$.
\end{itemize}
The proof can then be completed by summarizing the discussion above. $\hfill\blacksquare$


\subsection{Proof of the Results in Section~\ref{sec:sos_imp}}\label{proof_imp}
{\bf Proof of Theorem~\ref{thm:bilinear}.}
In general, we prove Theorem~\ref{thm:bilinear} by showing: 
1) ~\eqref{cd4}-\eqref{cdd1} imply (\textbf{Cond.1})-(\textbf{Cond.4}) in Definition~\ref{cond}, respectively;
2) ~\eqref{cd2} and \eqref{cd3} imply (\textbf{Cond.5}) in Definition~\ref{cond} holds, while~\eqref{cdd2} and~\eqref{cdd3} ensure that (\textbf{Cond.6}) in Definition~\ref{cond} are satisfied.

Firstly, one can prove~\eqref{cd4} $\Rightarrow$ (\textbf{Cond.1}),  and~\eqref{cdd4} $\Rightarrow$ (\textbf{Cond.3}) can be shown with the same ideas. 
If~\eqref{cd4} holds, then $\forall \mathsf{a}\in[1, n_{\mathsf{a}}]$,
one has $-\mathcal{B}(x)\geq\sum_{k=1}^{k_0(\mathsf{a})} s_{d,k}(\mathsf{a})\alpha_{\mathsf{a},k}(x)\geq 0$, $\forall x \in X_{0,\mathsf{a}}$.
The second inequality hold since $\forall k\in[1,k_0(\mathsf{a})]$, $s_{d,k}(\mathsf{a})\geq 0$ hold $\forall x \in \mathbb{R}^n$.
In other words, one has $ x\in X_0\Rightarrow x\in \mathcal{S}$, indicating that (\textbf{Cond.1}) holds.

Secondly, we show~\eqref{cd1} $\Rightarrow$ (\textbf{Cond.2}), while~\eqref{cdd1} $\Rightarrow$ (\textbf{Cond.4}), can be also be proved similarly.
One can verify (\textbf{Cond.2}) holds if $\forall \mathsf{c}\in[1,n_{\mathsf{c}}]$,
\begin{align}
	\big\{\gamma_{\mathsf{c},k}(x)\!\geq\! 0,\forall k\!\in\![1,k_d(\mathsf{c})], \mathcal{B}(x)\!\leq\! 0\big\}\!=\!\emptyset.\label{help1}
\end{align}
According to~\emph{Positivstellensatz}~\cite[Proposition 4.4.1]{Bochnak1998Real}, one has~\eqref{help1} holds if and only if there exists $s_{i_0,\ldots,i_{k_d(\mathsf{c})}}\in\mathcal{P}_S(x)$ such that
\begin{align*}
	\!\!\!\!\sum_{i_0,\ldots,i_{k_d(\mathsf{c})} \in \{0,1\}}\!\!\!\!\!\!\!\!\!\!\!\!\!\!s_{i_0,\ldots,i_{k_d(\mathsf{c})}}(-\mathcal{B}(x))^{i_0}\gamma_{\mathsf{c},1}^{i_1}(x)\cdots\gamma_{\mathsf{c},k_d(\mathsf{c})}^{i_{k_d(\mathsf{c})}}(x)\!\in\!\mathcal{P}_S(x).
\end{align*}
By selecting $s_{i_0,\ldots,i_{k_d(\mathsf{c})}}=0$ if $\sum_{r=1}^{k_d(\mathsf{c})}i_r>1$, one gets the relaxation as 
\begin{align}
	s - s_0 \mathcal{B}(x) + \sum_{k=1}^{k_d(\mathsf{c})}s_k\gamma_{\mathsf{c},k}(x)\in\mathcal{P}_S(x),\label{help4}
\end{align}
with $s,s_k\in\mathcal{P}_S(x)$, $\forall k\in[0,k_d(\mathsf{c})]$.
Note that~\eqref{help4} holds for all $\mathsf{c}\in[1,n_{\mathsf{c}}]$ if and only if~\eqref{cd1} holds.

	Finally, we show that~\eqref{cd2} and \eqref{cd3} imply (\textbf{Cond.5}) in Definition~\ref{cond} holds, while~\eqref{cdd2} and~\eqref{cdd3} ensure that (\textbf{Cond.6}) in Definition~\ref{cond} are satisfied.
	On one hand, one can verify that (\textbf{Cond.5}) and (\textbf{Cond.6}) in Definition~\ref{cond} hold if and only if~\eqref{cdd01}-\eqref{cd03} hold.
	On the other hand, if~\eqref{cddh01}-\eqref{cdh03} hold, then~\eqref{cdd01}-\eqref{cd03} hold.
	Here, we show that~\eqref{cd2}-\eqref{cdd3} implies~\eqref{cddh01}-\eqref{cdh03}, respectively.
	\begin{itemize}
		\item ~\eqref{cd2} $\Rightarrow$~\eqref{cddh01}: 
		With~\emph{Positivstellensatz}~\cite[Proposition 4.4.1]{Bochnak1998Real},
		~\eqref{cddh01} holds if and only if
		\begin{align}
			s_{b,0}-s_{b,1}\mathcal{B}(x)+&s_{b,2}\mathcal{B}(f(x,u))-s_{b,3}\mathcal{B}(x)\mathcal{B}(f(x,u))+\sum_{a=1}^{m}p_b(a)(u_a \!-\! k_a(x))\!=\!0 \label{hh1}
		\end{align}
		holds, with $s_{b,0},s_{b,1},s_{b,2},s_{b,3}\in\mathcal{P}_S(x,u)$ and $p_b(a)\in\mathcal{P}(x,u)$, $\forall a \in [1,m]$.
		By selecting $s_{b,3}=0$, one has~\eqref{hh1} holds if and only if~\eqref{cd2} holds. 
		\item ~\eqref{cd3} $\Rightarrow$~\eqref{cdh02}: 
		Similarly,~\eqref{cdh02} holds if and only if  $\forall j \in [1,\mathsf{j}]$, $\exists s_{c,i}(j) \in \mathcal{P}_S(x,u)$, with $i\in[0,3]$, $\exists  k\in \mathbb{N}$, and $\exists p_c(a,j)\in\mathcal{P}(x,u)$, with $a\in [1,m]$ such that
		\begin{align*}
			s_{c,0}(j) \!-\! s_{c,1}(j)&\mathcal{B}(x)\!+\!s_{c,2}(j)\rho_{j}(u)\!-\!s_{c,3}(j)\rho_{j}(u)\mathcal{B}(x)+\sum_{a=1}^{m}p_c(a,j)(u_a \!-\! \!k_a(x))\! +\!\rho_{j}(u)^{2k} \!=\!0,
		\end{align*}
		which holds if and only if~\eqref{cd3} holds.	
	\end{itemize}
	Then, the proof can be completed by showing~\eqref{cdd2}$\Rightarrow$\eqref{cdh01} and \eqref{cdd3}$\Rightarrow$\eqref{cdh03} in similar ways. $\hfill\blacksquare$
	
	{\bf Proof of Theorem~\ref{opt_nonlinear_init}}: 
	Firstly, according to~\cite[Lemma 4.1]{Seto1999Engineering}, one can verify
	$\{x\in \mathbb{R}^n|x^\top Q^{-1}x\leq 1\}\subseteq \bar{X}$ and
	$\{(x,\hat{x})\in \mathbb{R}^n\times \mathbb{R}^n|[x;\hat{x}]^\top Q_o^{-1}[x;\hat{x}]\leq 1\}\subseteq \bar{R}$,
	hold if and only if~\eqref{init_cond1-1} and~\eqref{init_cond2-1} hold, respectively.
	Therefore, one gets $\mathcal{S}\subseteq X\backslash X_d$ and $\mathcal{S}_O\subseteq \mathcal{R}\backslash \mathcal{R}_d$ with
	\begin{align}
		\mathcal{S}&:=\{x\in \mathbb{R}^n|x^\top Q^{-1}x\leq c_1\},\label{S}\\
		\mathcal{S}_O&:= \{(x,\hat{x})\in \mathbb{R}^n \times \mathbb{R}^n | [x;\hat{x}]^\top Q_o^{-1}[x;\hat{x}]\leq c_2 \},\label{S_o}
	\end{align}
	for all $c_1,c_2\in(0,1]$.
	Therefore, one has (\textbf{Cond.2}) and (\textbf{Cond.4}) hold with $\mathcal{S}$ and $\mathcal{S}_O$ as in~\eqref{S} and~\eqref{S_o}.
	
	Finally, we show that~\eqref{init_cond1} and~\eqref{init_cond2} imply (\textbf{Cond.5}) and (\textbf{Cond.6}), respectively.
	\begin{itemize}
		\item ~\eqref{init_cond1} $\Rightarrow$ (\textbf{Cond.5}): 
		We show~\eqref{init_cond1} implies that there exists $c_1\in(0,1]$ such that (\textbf{Cond.5}) holds with $\mathcal{S}$ as in~\eqref{S}, when $u(x):=\bar{K}(x)Q^{-1}x$ is applied for all $x\in\mathcal{S}$.
		Consider a controller $u(x):=K(x)x$, with $K(x)\in\mathcal{P}(x)$.
		For all $x\in \mathcal{S}$, one has $A\mathcal{H}(x)x+B\mathcal{U}(x)u\in\mathcal{S}$ if
		$P - (A\mathcal{H}(x)+B\mathcal{U}(x)K(x))^{\top}\!P(A\mathcal{H}(x)+B\mathcal{U}(x)K(x))\in \mathcal{P}_S^{\mathsf{m}}(x)$
		holds, with $P=Q^{-1}$, or, equivalently, 
		\begin{align}
			Q - (A&\mathcal{H}(x)Q+B\mathcal{U}(x)\bar{K}(x))^{\top}P(A\mathcal{H}(x)Q+B\mathcal{U}(x)\bar{K}(x))\in \mathcal{P}_S^{\mathsf{m}}(x),\label{help2}
		\end{align}
		holds, with $\bar{K}(x)=K(x)Q$.
		Then, considering the Schur complement~\cite{Zhang2006Schur} of $Q$ of the matrix on the left-hand-side of~\eqref{init_cond1}, one has~\eqref{help2} holds if~\eqref{init_cond1} holds.
		Since $0_n\in\mathcal{S}$, and $u(x)=K(x)x=0$ when $x=0_n$, there must exists $c_1\in\mathbb{R}_{>0}$ such that $u(x)\in U$ holds $\forall x \in \mathcal{S}$.
		Hence, one has~\eqref{init_cond1} $\Rightarrow$ (\textbf{Cond.5}).
		\item ~\eqref{init_cond2} $\Rightarrow$ (\textbf{Cond.6}): 
		Next, we show~\eqref{init_cond2} implies that there exist $c_2\in(0,1]$ such that (\textbf{Cond.6}) holds with $\mathcal{S}_O$ as in~\eqref{S_o}, when $u(x):=\bar{K}(x)Q^{-1}x$ and $\hat{u}(x,\hat{x}):=\bar{K}_o(x,\hat{x})Q_o^{-1}[x;\hat{x}]$ are applied for all $[x;\hat{x}]\in\mathcal{S}_O$.
		According to~\cite[Theorem 1.12]{Zhang2006Schur}, if~\eqref{init_cond2} holds, one has
		$Q_o -(A_o(x,\hat{x})Q_o+B_o(\hat{x})\bar{K}'_o(x,\hat{x}))^{\top}Q_o^{-1}(A_o(x,\hat{x})Q_o+B_o(\hat{x})\bar{K}'_o(x,\hat{x}))\in \mathcal{P}_S^{\mathsf{m}}(x,\hat{x})$,
		which is equivalent to
		\begin{align}
			&Q^{-1}_o - (A_o(x,\hat{x})+B_o(\hat{x})K'_o(x,\hat{x}))^{\top}Q_o^{-1}(A_o(x,\hat{x})+B_o(\hat{x})K'_o(x,\hat{x}))\in \mathcal{P}_S^{\mathsf{m}}(x,\hat{x}),\label{help6}
		\end{align}
		with $K'_o(x,\hat{x}):=\bar{K}'_o(x,\hat{x})Q^{-1}_o$.
		On one hand, if~\eqref{help6} holds, then given any $c_2\in(0,1]$, for all $[x;\hat{x}]\in\mathbb{R}^{2n}$ satisfying $[x;\hat{x}]^\top Q^{-1}_o[x;\hat{x}]\leq c_2$, one has 
		$[x^+;\hat{x}^+]^\top Q^{-1}_o[x^+;\hat{x}^+]\leq c_2$,
		in which $x^+=A\mathcal{H}(x)x+B\mathcal{U}(x)\bar{K}(x)Q^{-1}x$ and $\hat{x}^+=A\mathcal{H}(\hat{x})\hat{x}+B\mathcal{U}(\hat{x})\bar{K}_o(x,\hat{x})Q_o^{-1}\hat{x}$.
		On the other hand, similar to the previous case, there exists $c_2\in\mathbb{R}_{>0}$ such that $\hat{u}(x,\hat{x})\in U$ since $0_{2n}\in\mathcal{S}_O$.
		Therefore, one can conclude that~\eqref{init_cond2} implies that (\textbf{Cond.6}).
	\end{itemize}

\subsection{Omitted Figures, Algorithms, and Mathematical Expressions in the Main Text}\label{ap3}
\begin{figure}
	\begin{center}
		\includegraphics[width=0.5\textwidth]{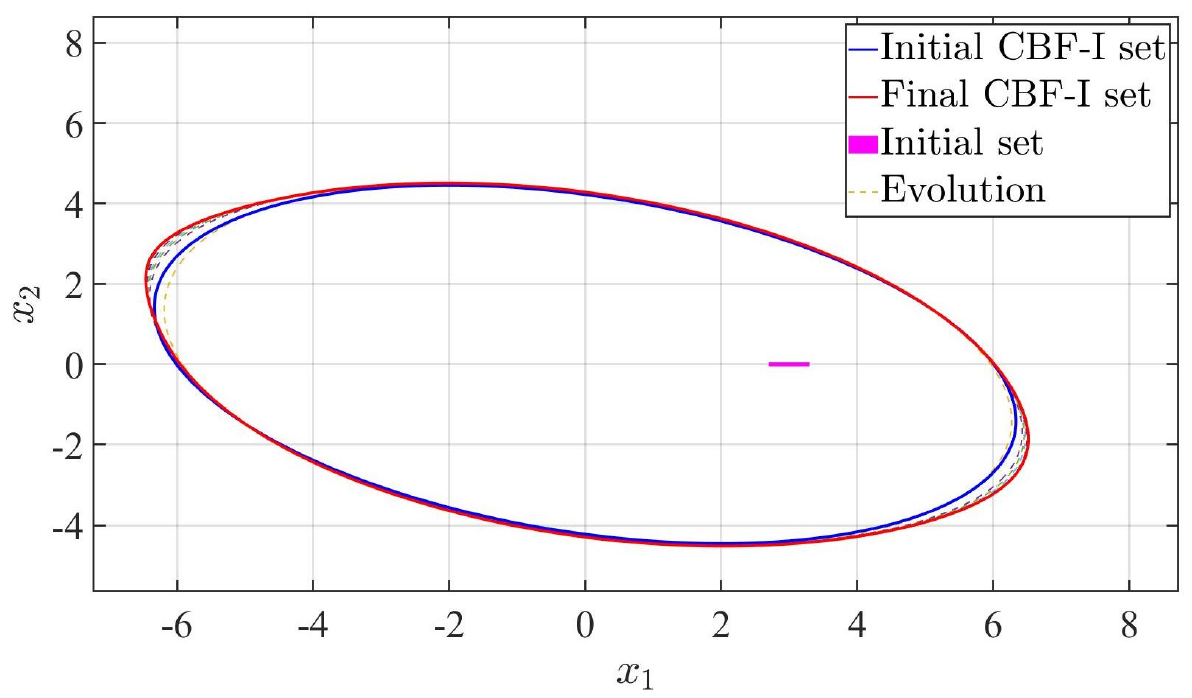}    
		\caption{Evolution of the CBF-I set with respect to the number of iterations for the running example.} 
		\label{fig:CBF}
	\end{center}
\end{figure}

\begin{figure}
	\begin{center}
		\includegraphics[width=0.45\textwidth]{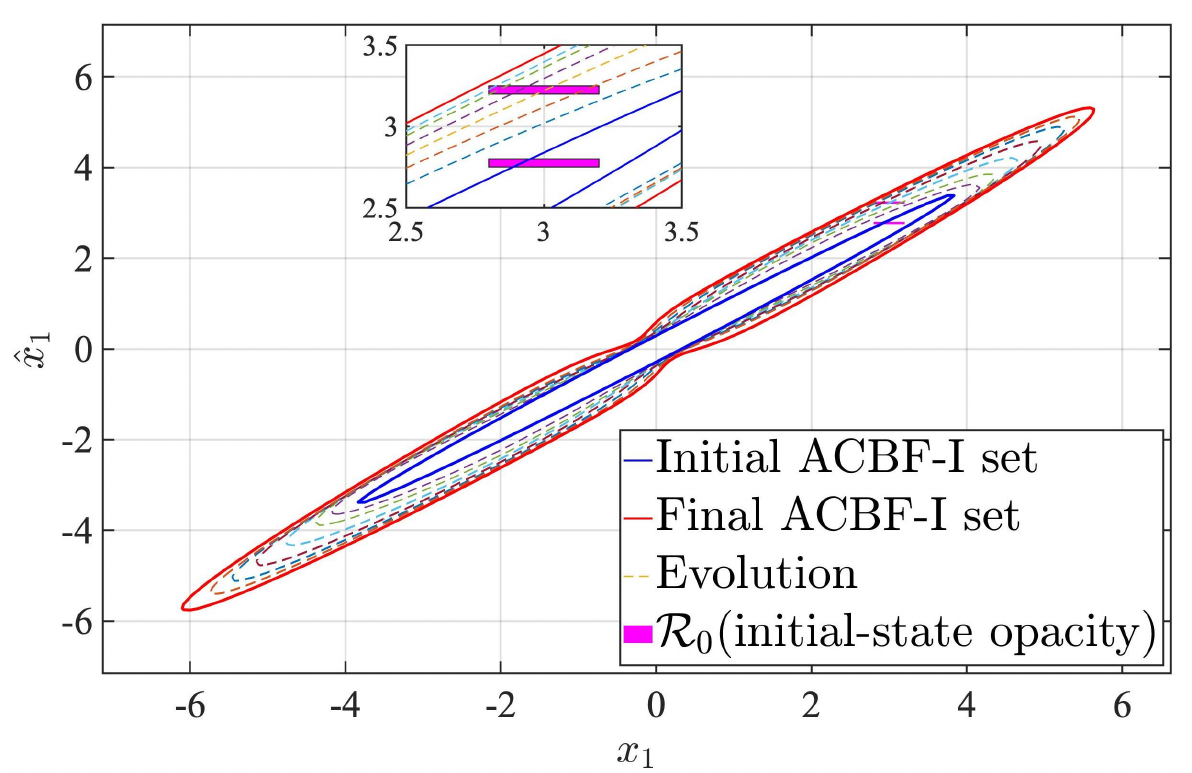}    
		\caption{Evolution of the cross-section of the ACBF-I set with respect to the number of iterations for the running example, with $x_2=\hat{x}_2=0$, and $\mathcal{R}_0$ as in~\eqref{R0init}.} 
		\label{fig:ACBF}
	\end{center}
\end{figure}

In this subsection, we propose some of the figures, algorithms, and mathematical expressions that are omitted in the main text of the paper.
Concretely, Algorithm~\ref{alg:initial} can be used to compute $\mathcal{S}^{\text{init}}$ and $\mathcal{S}^{\text{init}}_O$ systematically based on Theorem~\ref{opt_nonlinear_init}.
Moreover, concerning the case study of the running example in Section~\ref{cases:car}, the evolution of the sets $\mathcal{S}^{\text{exp}}$ and $\mathcal{S}_O^{\text{exp}}$ with respect to the number of iterations are depicted in Figure~\ref{fig:CBF} and Figure~\ref{fig:ACBF}, respectively.
\IncMargin{0.5em}
\begin{algorithm2e}[ht!]
	\DontPrintSemicolon
	\Indm 
	\KwIn{System as in~\eqref{sys2}, sets $\bar{X}$ and $\bar{\mathcal{R}}$ as in~\eqref{X_sub} and \eqref{R_sub}, respectively.
		Maximal degrees of the matrix polynomials $\bar{K}(x)$ and $\bar{K}_o(x,\hat{x})$ in Theorem \ref{opt_nonlinear_init}, denoted by $d_{\bar{k}}^{\text{max}}$ and $d_{\bar{k}_o}^{\text{max}}$, respectively.}
	\KwOut{Initial CBF-I set $\mathcal{S}^{\text{init}}= \{x\in \mathbb{R}^n | x^\top Q^{-1}x -c_1\leq 0  \}$ and the initial ACBF-I set $\mathcal{S}_O^{\text{init}} = \{(x,\hat{x})\in \mathbb{R}^n \times \mathbb{R}^n | [x;\hat{x}]^\top Q_o^{-1}[x;\hat{x}]- c_2\leq 0\}$; Otherwise, the algorithm stops inconclusively if one cannot find $\bar{K}(x)$, $Q$, $\bar{K}_o(x,\hat{x})$ or $Q_o$ for the given $d_{\bar{k}}^{\text{max}}$ and $d_{\bar{k}_o}^{\text{max}}$.}
	\Indp	
	Compute $\bar{K}(x)$ and $Q$ in~\eqref{init_cond1} and~\eqref{init_cond1-1}.
	If $\bar{K}(x)$ and $Q$ are obtained successfully, proceed to step~\ref{step_2}; otherwise, increase the degree of $\bar{K}(x)$ (up to $d_{\bar{k}}^{\text{max}}$) and repeat 
	step~\ref{step_1}.\label{step_1}\\
	Set $K(x):=\bar{K}(x)Q^{-1}$ with $\bar{K}(x)$ and $Q$ obtained in step~\ref{step_1}. 
	Compute $\bar{K}_o(x,\hat{x})$ and $Q_o$ in~\eqref{init_cond2} and~\eqref{init_cond2-1}.
	Increase the degree of $\bar{K}_o(x,\hat{x})$ (up to $d_{\bar{k}_o}^{\text{max}}$) and repeat step~\ref{step_2} if  $\bar{K}_o(x,\hat{x})$ and $Q_o$ cannot be found;
	otherwise, proceed to step~\ref{step_3}.\label{step_2} \\
	Compute $c_1$ with bisection over $c_1\in (0,1]$ such that~\eqref{cd3} holds with $u(x) = \bar{K}(x)Q^{-1}x$ and $\mathcal{B}(x) = x^\top Q^{-1}x - c_1$, in which $\bar{K}(x)$ and $Q$ are obtained in step~\ref{step_1}.\label{step_3}\\
	Compute $c_2$ with bisection over $c_2\in (0,1]$, such that~\eqref{cdd3} holds with $\hat{u}=\bar{K}_o(x,\hat{x})Q_o^{-1}[x;\hat{x}]$, and $\mathcal{B}_O(x,\hat{x})= [x;\hat{x}]^\top Q^{-1}_o [x;\hat{x}]- c_2$,  in which $Q_o$ and $\bar{K}_o(x,\hat{x})$ are obtained in step~\ref{step_2}.\label{step_4}\\
	\caption{Computing initial CBF-I set $\mathcal{S}^{\text{init}}$ and initial ACBF-I set $\mathcal{S}_O^{\text{init}}$ for systems as in~\eqref{sys2}.}
	\label{alg:initial}
\end{algorithm2e}
\DecMargin{0.5em}
Finally, we summarize those expressions for computing the initial (resp. expanded) CBF-I set $\mathcal{S}^{\text{init}}$ (resp. $\mathcal{S}^{\text{exp}}$) and the initial (resp. expanded) ACBF-I set $\mathcal{S}_O^{\text{init}}$ (resp. $\mathcal{S}_O^{\text{exp}}$) for the case studies of this paper. 
Note that those parameters that are smaller than $10^{-4}$ are neglected for simple presentation.

{\bf Running example}
The CBF $\mathcal{B}(x)$, ACBF $\mathcal{B}_O(x,\hat{x})$, functions $k(x)$ and $\hat{k}(x,\hat{x})$ associated with the expanded $\mathcal{S}^{\text{exp}}$ and $\mathcal{S}_O^{\text{exp}}$ for the running example are as follows.
\begin{align*}
	\mathcal{B}(x)&=0.5688-0.6237 x_1+0.2971 x_2-0.6125 x_1^2-1.7204 x_1 x_2-2.4687 x_2^2+0.0162 x_1^3+0.0093 x_1^2 x_2\nonumber\\
	&+0.0296 x_1 x_2^2-0.0145 x_2^3+0.0174 x_1^4+0.0649 x_1^3 x_2+0.1433 x_1^2 x_2^2+0.1533 x_1 x_2^3+0.1353 x_2^4;\\
	\mathcal{B}_O(x,\hat{x})&=0.5633-0.0006 x_1-0.0012 x_2-0.0018 \hat{x}_1-0.0002 \hat{x}_2+0.0138 x_1^2-0.0201 x_1 x_2-0.0413 x_2^2\nonumber\\
	&-0.1231 x_1 \hat{x}_1-0.0482 x_2 \hat{x}_1+0.0029 \hat{x}_1^2-0.0278 x_1 \hat{x}_2-0.045 x_2 \hat{x}_2-0.0061 \hat{x}_1 \hat{x}_2-0.0065 \hat{x}_2^2\nonumber\\
	&+0.001 x_1^3+0.0012 x_1^2 x_2-0.0014 x_1 x_2^2-0.0011 x_1^2 \hat{x}_1-0.0013 x_1 x_2 \hat{x}_1+0.0012 x_2^2 \hat{x}_1-0.0012 x_1 \hat{x}_1^2\nonumber\\
	&+0.0001 x_2 \hat{x}_1^2+0.0031 \hat{x}_1^3-0.0004 x_1^2 \hat{x}_2+0.0005 x_1 x_2 \hat{x}_2+0.0001 x_2^2 \hat{x}_2-0.0001 x_1 \hat{x}_1 \hat{x}_2\nonumber\\
	&+0.0013 \hat{x}_1^2 \hat{x}_2+0.0007 x_1 \hat{x}_2^2+0.0004 x_2 \hat{x}_2^2-0.0004 \hat{x}_1 \hat{x}_2^2-0.0004 \hat{x}_2^3+0.0349 x_1^4+0.0515 x_1^3 x_2\nonumber\\
	&+0.0511 x_1^2 x_2^2+0.0151 x_1 x_2^3+0.0215 x_2^4+0.001 x_1^3 \hat{x}_1-0.0153 x_1^2 x_2 \hat{x}_1-0.0514 x_1 x_2^2 \hat{x}_1+0.0135 x_2^3 \hat{x}_1\nonumber\\
	&-0.0699 x_1^2 \hat{x}_1^2-0.0707 x_1 x_2 \hat{x}_1^2+0.0434 x_2^2 \hat{x}_1^2-0.0145 x_1 \hat{x}_1^3+0.0338 x_2 \hat{x}_1^3+0.0523 \hat{x}_1^4+0.0112 x_1^3 \hat{x}_2\nonumber\\
	&+0.0491 x_1^2 x_2 \hat{x}_2-0.0037 x_1 x_2^2 \hat{x}_2-0.0133 x_2^3 \hat{x}_2-0.0392 x_1^2 \hat{x}_1 \hat{x}_2-0.1664 x_1 x_2 \hat{x}_1 \hat{x}_2-0.0284 x_2^2 \hat{x}_1 \hat{x}_2\nonumber\\
	&-0.011 x_1 \hat{x}_1^2 \hat{x}_2+0.049 x_2 \hat{x}_1^2 \hat{x}_2+0.0455 \hat{x}_1^3 \hat{x}_2+0.0269 x_1^2 \hat{x}_2^2-0.0066 x_1 x_2 \hat{x}_2^2-0.0164 x_2^2 \hat{x}_2^2+0.011 \hat{x}_2^4\nonumber\\
	&-0.0316 x_1 \hat{x}_1 \hat{x}_2^2-0.0032 x_2 \hat{x}_1 \hat{x}_2^2+0.0404 \hat{x}_1^2 \hat{x}_2^2-0.0043 x_1 \hat{x}_2^3+0.0005 x_2 \hat{x}_2^3+0.0235 \hat{x}_1 \hat{x}_2^3;\\
	k(x)&=-0.004-0.5205 x_1-0.9476 x_2+0.0075 x_1^2+0.1525 x_1 x_2+0.0872 x_2^2+0.0429 x_1^3+0.1126 x_1^2 x_2\nonumber\\
	&+0.0852 x_1 x_2^2+0.12 x_2^3+0.0003 x_1^4-0.0134 x_1^3 x_2-0.0274 x_1^2 x_2^2-0.0287 x_1 x_2^3-0.0137 x_2^4-0.0012 x_1^5\nonumber\\
	&-0.0042 x_1^4 x_2-0.0068 x_1^3 x_2^2-0.0109 x_1^2 x_2^3-0.0065 x_1 x_2^4-0.0064 x_2^5+0.0002 x_1^5 x_2+0.0008 x_1^4 x_2^2\nonumber\\
	&+0.0015 x_1^3 x_2^3+0.0018 x_1^2 x_2^4+0.0012 x_1 x_2^5+0.0005 x_2^6;\\
	\hat{k}(x,\hat{x})&=0.0033+1.9794 x_1+1.3667 x_2-1.9404 \hat{x}_1-1.9815 \hat{x}_2-0.0613 x_1^2+0.1014 x_1 x_2+0.076 x_2^2\nonumber\\
	&-0.119 x_1 \hat{x}_1-0.0608 x_2 \hat{x}_1+0.1039 \hat{x}_1^2-0.0119 x_1 \hat{x}_2-0.1312 x_2 \hat{x}_2-0.0957 \hat{x}_1 \hat{x}_2+0.0283 \hat{x}_2^2\nonumber\\
	&-0.177 x_1^3-0.2866 x_1^2 x_2-0.3162 x_1 x_2^2-0.1003 x_2^3+0.2124 x_1^2 \hat{x}_1+0.4715 x_1 x_2 \hat{x}_1+0.1037 x_2^2 \hat{x}_1\nonumber\\
	&+0.1796 x_1 \hat{x}_1^2-0.1623 x_2 \hat{x}_1^2-0.2304 \hat{x}_1^3+0.1444 x_1^2 \hat{x}_2+0.5553 x_1 x_2 \hat{x}_2+0.2728 x_2^2 \hat{x}_2-0.1237 x_1 \hat{x}_1 \hat{x}_2\nonumber\\
	&-0.2061 x_2 \hat{x}_1 \hat{x}_2-0.0491 \hat{x}_1^2 \hat{x}_2-0.2777 x_1 \hat{x}_2^2-0.2637 x_2 \hat{x}_2^2+0.1323 \hat{x}_1 \hat{x}_2^2+0.0954 \hat{x}_2^3+0.0235 x_1^4\nonumber\\
	&+0.0253 x_1^3 x_2+0.0753 x_1^2 x_2^2+0.0339 x_1 x_2^3+0.0398 x_2^4-0.0116 x_1^3 \hat{x}_1+0.0219 x_1^2 x_2 \hat{x}_1-0.0542 x_1 x_2^2 \hat{x}_1\nonumber\\
	&-0.0198 x_2^3 \hat{x}_1+0.0192 x_1^2 \hat{x}_1^2-0.1257 x_1 x_2 \hat{x}_1^2-0.0177 x_2^2 \hat{x}_1^2-0.0944 x_1 \hat{x}_1^3+0.0786 x_2 \hat{x}_1^3+0.0656 \hat{x}_1^4\nonumber\\
	&-0.0244 x_1^3 \hat{x}_2-0.1104 x_1^2 x_2 \hat{x}_2-0.0848 x_1 x_2^2 \hat{x}_2-0.1286 x_2^3 \hat{x}_2+0.0306 x_1^2 \hat{x}_1 \hat{x}_2+0.0554 x_1 x_2 \hat{x}_1 \hat{x}_2\nonumber\\
	&+0.0707 x_2^2 \hat{x}_1 \hat{x}_2+0.0178 x_1 \hat{x}_1^2 \hat{x}_2+0.0453 x_2 \hat{x}_1^2 \hat{x}_2-0.0205 \hat{x}_1^3 \hat{x}_2+0.0536 x_1^2 \hat{x}_2^2+0.0381 x_1 x_2 \hat{x}_2^2\nonumber\\
	&+0.157 x_2^2 \hat{x}_2^2-0.028 x_1 \hat{x}_1 \hat{x}_2^2-0.0535 x_2 \hat{x}_1 \hat{x}_2^2-0.0142 \hat{x}_1^2 \hat{x}_2^2+0.0032 x_1 \hat{x}_2^3-0.0897 x_2 \hat{x}_2^3\nonumber\\
	&+0.0149 \hat{x}_1 \hat{x}_2^3+0.022 \hat{x}_2^4.\\
\end{align*}

{\bf Case study of chaser satellite}
After computing the initial ACBF-I set $\mathcal{S}_O^{\text{init}}$ for the chaser satellite case study, matrices $\bar{K}$ and $Q_0$ are as follows:
\begin{align*}
		\bar{K}_o &= \begin{bmatrix}
		0.136 &   -0.003 &   0.215 &   -0.274 &   -0.146 &   0 &   -0.214 &    0.220\\
		2.704 &   -0.657 &    5.223 &   -8.853 &   -4.164 &   -0.001 &   -5.476 &    0.163
	\end{bmatrix},\\
	Q_o &= \begin{bmatrix}
		654.4078 &	-58.657&	-6.321&	  -102.935&	 649.790&	-62.536&	-6.598&	 -103.187\\
		  -58.657&	 97.155&	-28.654&	 2.918&	 -57.566& 	 94.343&	-28.670&	3.177\\
		   -6.321&	-28.654&	 92.077&   -0.0244&	 -6.136&	-34.657&	 92.152&	-0.007\\
		 -102.935&	  2.918&	-0.0244&	17.443&	-102.570&	  3.354&	-0.0423&	17.457\\
		  649.790&	-57.566&	-6.136&	  -102.570&	 647.025&	-121.345&	 -6.853&	-102.960\\
		  -62.536&	 94.343&	-34.657&	 3.354&	-121.345&	198380.007&	-28.577&	3.937\\
		   -6.598&	-28.670&	 92.152&	-0.043&	  -6.853&	-28.5774&	94.052&	-0.028\\
		 -103.187&	  3.177&	-0.007&	    17.457&	-102.960&	  3.937&	-0.028&	17.833
	\end{bmatrix}.
\end{align*}

The CBF $\mathcal{B}(x)$, ACBF $\mathcal{B}_O(x,\hat{x})$, functions $k_1(x)$, $k_2(x)$, $\hat{k}_1(x,\hat{x})$, and $\hat{k}_2(x,\hat{x})$ associated with the expanded $\mathcal{S}^{\text{exp}}$ and $\mathcal{S}_O^{\text{exp}}$ for the chaser satellite case study are as follows:
\begin{align*}
	\mathcal{B}(x)&=0.9978+0.0068 x_1+0.006 x_2+0.0101 x_3+0.0505 x_4-0.0566 x_1^2-0.007 x_1 x_2-0.0271 x_2^2\nonumber\\
	&-0.0297 x_1 x_3-0.0285 x_2 x_3-0.152 x_3^2-0.5124 x_1 x_4-0.0041 x_2 x_4-0.2534 x_3 x_4-1.4303 x_4^2\nonumber\\
	&-0.0002 x_1^3-0.0005 x_1^2 x_2-0.0006 x_1 x_2^2-0.0002 x_2^3-0.0003 x_1^2 x_3-0.0003 x_1 x_2 x_3-0.0005 x_2^2 x_3\nonumber\\
	&-0.0001 x_2 x_3^2-0.0001 x_3^3-0.0043 x_1^2 x_4-0.0061 x_1 x_2 x_4-0.0035 x_2^2 x_4-0.0038 x_1 x_3 x_4-0.0021 x_2 x_3 x_4\nonumber\\
	&-0.0005 x_3^2 x_4-0.026 x_1 x_4^2-0.0175 x_2 x_4^2-0.0114 x_3 x_4^2-0.0533 x_4^3+0.0017 x_1^4+0.0025 x_1^3 x_2\nonumber\\
	&+0.0032 x_1^2 x_2^2+0.0004 x_1 x_2^3+0.0012 x_2^4+0.0009 x_1^3 x_3+0.0019 x_1^2 x_2 x_3+0.0017 x_1 x_2^2 x_3+0.001 x_2^3 x_3\nonumber\\
	&+0.0054 x_1^2 x_3^2+0.0021 x_1 x_2 x_3^2+0.0056 x_2^2 x_3^2+0.001 x_1 x_3^3+0.0006 x_2 x_3^3+0.0011 x_3^4+0.0355 x_1^3 x_4\nonumber\\
	&+0.039 x_1^2 x_2 x_4+0.0299 x_1 x_2^2 x_4+0.0011 x_2^3 x_4+0.0181 x_1^2 x_3 x_4+0.0213 x_1 x_2 x_3 x_4+0.0127 x_2^2 x_3 x_4\nonumber\\
	&+0.0618 x_1 x_3^2 x_4+0.012 x_2 x_3^2 x_4+0.0063 x_3^3 x_4+0.2807 x_1^2 x_4^2+0.2065 x_1 x_2 x_4^2+0.0813 x_2^2 x_4^2\nonumber\\
	&+0.1209 x_1 x_3 x_4^2+0.0622 x_2 x_3 x_4^2+0.1801 x_3^2 x_4^2+1.0191 x_1 x_4^3+0.3767 x_2 x_4^3+0.2687 x_3 x_4^3+1.4473 x_4^4;\\
	\mathcal{B}_O(x,\hat{x})&= -0.0019+0.0002 x_1+0.0004 x_2-0.002 x_3+0.0094 x_4+0.0009  \hat{x}_1+0.0027 \hat{x}_3-0.0036\hat{x}_4\nonumber\\
	&+0.2636 x_1^2-0.0107 x_1 x_2+0.0107 x_2^2+0.0361 x_1 x_3-0.0312 x_2 x_3+0.2944 x_3^2+0.0113 x_1 x_4\nonumber\\
	&+0.1022 x_2 x_4-0.1939 x_3 x_4+0.9 x_4^2-0.5482 x_1 \hat{x}_1+0.0389 x_2 \hat{x}_1-0.0876 x_3  \hat{x}_1+0.1269 x_4  \hat{x}_1\nonumber\\
	&+0.3112  \hat{x}_1^2+0.0002 x_1  \hat{x}_2-0.0005 x_2  \hat{x}_2+0.0009 x_3  \hat{x}_2+0.0045 x_4  \hat{x}_2-0.0005  \hat{x}_1  \hat{x}_2+0.0003  \hat{x}_2^2\nonumber\\
	&-0.0307 x_1  \hat{x}_3+0.0333 x_2  \hat{x}_3-0.59 x_3  \hat{x}_3+0.2022 x_4  \hat{x}_3+0.0843  \hat{x}_1  \hat{x}_3-0.001  \hat{x}_2  \hat{x}_3+0.2991  \hat{x}_3^2\nonumber\\
	&-0.1601 x_1  \hat{x}_4+0.0492 x_2  \hat{x}_4-0.1228 x_3  \hat{x}_4-1.2124 x_4  \hat{x}_4+0.3105  \hat{x}_1  \hat{x}_4-0.0068  \hat{x}_2  \hat{x}_4\nonumber\\
	&+0.1209  \hat{x}_3  \hat{x}_4+1.1352  \hat{x}_4^2;\\
	k_1(x)  &= 0.0014-0.6836 x_1-0.7086 x_2-0.3108 x_3-2.3368 x_4+0.0003 x_1^2+0.0006 x_1 x_2+0.0004 x_2^2\nonumber\\
	&-0.0001 x_1 x_3+0.0003 x_3^2+0.0032 x_1 x_4+0.0039 x_2 x_4+0.0003 x_3 x_4+0.0118 x_4^2;\\
	k_2(x)  &= 	0.0027-1.7315 x_1-1.2185 x_2-0.0702 x_3-10.0234 x_4+0.0003 x_1^2+0.0004 x_1 x_2+0.0007 x_2^2\nonumber\\
	&+0.0009 x_2 x_3+0.0001 x_3^2+0.0025 x_1 x_4+0.0023 x_2 x_4-0.0004 x_3 x_4+0.0066 x_4^2;\\
	\hat{k}_1(x,\hat{x}) &= 	0.0264+0.0026 x_1-0.0049 x_2+0.1314 x_3-0.0478 x_4+0.0064  \hat{x}_1-0.0107  \hat{x}_2-0.1446  \hat{x}_3\nonumber\\
	&+0.0901  \hat{x}_4+0.0234 x_1^2-0.0027 x_1 x_2-0.0019 x_2^2+0.0151 x_1 x_3+0.0058 x_2 x_3-0.0173 x_3^2\nonumber\\
	&-0.0323 x_1 x_4-0.0301 x_2 x_4+0.0331 x_3 x_4-0.1815 x_4^2-0.0554 x_1  \hat{x}_1-0.0016 x_2  \hat{x}_1-0.0102 x_3  \hat{x}_1\nonumber\\
	&-0.0054 x_4  \hat{x}_1+0.0289  \hat{x}_1^2-0.0002 x_1  \hat{x}_2-0.0002 x_2  \hat{x}_2+0.0004 x_3  \hat{x}_2-0.0006 x_4  \hat{x}_2+0.0002  \hat{x}_1  \hat{x}_2\nonumber\\
	&-0.0156 x_1  \hat{x}_3-0.005 x_2  \hat{x}_3+0.0328 x_3  \hat{x}_3-0.0306 x_4  \hat{x}_3+0.0122  \hat{x}_1  \hat{x}_3-0.0005  \hat{x}_2  \hat{x}_3-0.016  \hat{x}_3^2\nonumber\\
	&-0.0012 x_1  \hat{x}_4+0.0073 x_2  \hat{x}_4+0.0056 x_3  \hat{x}_4+0.1384 x_4  \hat{x}_4+0.0084  \hat{x}_1  \hat{x}_4+0.0006  \hat{x}_2  \hat{x}_4\nonumber\\
	&-0.0005  \hat{x}_3  \hat{x}_4-0.0611  \hat{x}_4^2;\\
	\hat{k}_2(x,\hat{x}) &= 	-0.0036+0.1768 x_1-0.7756 x_2+3.6439 x_3-4.5077 x_4-1.5763  \hat{x}_1+0.0078  \hat{x}_2-3.63  \hat{x}_3\nonumber\\
	&-3.521  \hat{x}_4-0.0001 x_1^2-0.001 x_1 x_2+0.0006 x_2^2-0.0149 x_1 x_3+0.0014 x_2 x_3-0.0103 x_3^2\nonumber\\
	&-0.0041 x_1 x_4+0.008 x_2 x_4-0.0005 x_3 x_4+0.0329 x_4^2+0.0051 x_1  \hat{x}_1+0.0027 x_2  \hat{x}_1+0.0315 x_3  \hat{x}_1\nonumber\\
	&+0.03 x_4  \hat{x}_1-0.0037  \hat{x}_1^2+0.001 x_1  \hat{x}_2+0.0001 x_2  \hat{x}_2+0.0008 x_3  \hat{x}_2+0.0012 x_4  \hat{x}_2-0.0009  \hat{x}_1  \hat{x}_2\nonumber\\
	&+0.0001  \hat{x}_2^2+0.0179 x_1  \hat{x}_3+0.0005 x_2  \hat{x}_3+0.0265 x_3  \hat{x}_3+0.0125 x_4  \hat{x}_3-0.0303  \hat{x}_1  \hat{x}_3-0.0007  \hat{x}_2  \hat{x}_3\nonumber\\
	&-0.0152  \hat{x}_3^2-0.0017 x_1  \hat{x}_4-0.0058 x_2  \hat{x}_4+0.0382 x_3  \hat{x}_4-0.0308 x_4  \hat{x}_4-0.018  \hat{x}_1  \hat{x}_4-0.0006  \hat{x}_2  \hat{x}_4\nonumber\\
	&-0.0413  \hat{x}_3  \hat{x}_4-0.0077  \hat{x}_4^2.\\
\end{align*}

\end{document}